\documentclass[sigconf]{acmart}

\AtBeginDocument{%
  }

\copyrightyear{2025}
\acmYear{2025}
\setcopyright{cc}
\setcctype{by}
\acmConference[SIGIR '25]{Proceedings of the 48th International ACM SIGIR Conference on Research and Development in Information Retrieval}{July 13--18, 2025}{Padua, Italy}
\acmBooktitle{Proceedings of the 48th International ACM SIGIR Conference on Research and Development in Information Retrieval (SIGIR '25), July 13--18, 2025, Padua, Italy}\acmDOI{10.1145/3726302.3729953}
\acmISBN{979-8-4007-1592-1/2025/07}
\usepackage{subcaption}
\usepackage{soul}
\usepackage{booktabs}
\usepackage{multirow}
\usepackage{amsmath}
\usepackage{enumitem}
\usepackage{pifont}
\usepackage{xcolor}
\usepackage{hyperref}

\newcommand{\mm}[1]{\(#1\)}
\newcommand{\mc}[1]{\mathcal{#1}}
\newcommand{\mb}[1]{\mathbb{#1}}
\newcommand{\proposed}{\textsf{DGMRec}}

\newcommand{\cmark}{\textcolor{green!80!black}{\ding{51}}}
\newcommand{\xmark}{\textcolor{red}{\ding{55}}}

\begin{document}

\title{Disentangling and Generating Modalities for Recommendation in Missing Modality Scenarios}

\author{Jiwan Kim}
\email{kim.jiwan@kaist.ac.kr}
\affiliation{%
  \institution{KAIST}
  \city{Daejeon}
  \country{Republic of Korea}
}

\author{Hongseok Kang}
\email{ghdtjr0311@kaist.ac.kr}
\affiliation{%
  \institution{KAIST}
  \city{Daejeon}
  \country{Republic of Korea}
}

\author{Sein Kim}
\email{rlatpdlsgns@kaist.ac.kr}
\affiliation{%
  \institution{KAIST}
  \city{Daejeon}
  \country{Republic of Korea}
}

\author{Kibum Kim}
\email{kb.kim@kaist.ac.kr}
\affiliation{%
  \institution{KAIST}
  \city{Daejeon}
  \country{Republic of Korea}
}

\author{Chanyoung Park}
\authornote{Corresponding author}
\email{cy.park@kaist.ac.kr}
\affiliation{%
  \institution{KAIST}
  \city{Daejeon}
  \country{Republic of Korea}
}
\renewcommand{\shortauthors}{Jiwan Kim, Hongseok Kang, Sein Kim, Kibum Kim, \& Chanyoung Park}

\begin{abstract}
Multi-modal recommender systems (MRSs) have achieved notable success in improving personalization by leveraging diverse modalities such as images, text, and audio. However, two key challenges remain insufficiently addressed: (1) Insufficient consideration of missing modality scenarios and (2) the overlooking of unique characteristics of modality features. These challenges result in significant performance degradation in realistic situations where modalities are missing.
To address these issues, we propose \textbf{D}isentangling and \textbf{G}enerating \textbf{M}odality \textbf{Rec}ommender (\proposed), a novel framework tailored for missing modality scenarios. \proposed~ disentangles modality features into general and specific modality features from an information-based perspective, enabling richer representations for recommendation. Building on this, it generates missing modality features by integrating aligned features from other modalities and leveraging user modality preferences.
Extensive experiments show that \proposed~ consistently outperforms state-of-the-art MRSs in challenging scenarios, including missing modalities and new item settings as well as diverse missing ratios and varying levels of missing modalities. Moreover, \proposed's generation-based approach enables cross-modal retrieval, a task inapplicable for existing MRSs, highlighting its adaptability and potential for real-world applications.  Our code is available at \href{https://github.com/ptkjw1997/DGMRec}{https://github.com/ptkjw1997/DGMRec}.

\end{abstract}



\maketitle

\section{Introduction}
In recent years, e-commerce platforms such as Amazon and Alibaba, as well as social media services like YouTube and TikTok, have become an integral part of everyday life.
Their recommendation systems play a pivotal role in shaping user behavior and decision-making. 
In particular, traditional methods using Collaborative Filtering (CF), such as Matrix Factorization (MF) \cite{rendle2012bpr} or Graph Neural Networks (GNNs) \cite{he2020lightgcn, wang2019ngcf}, have successfully delivered personalized recommendations, making these systems increasingly important for both customers and businesses.

\begin{figure}[t!]
    \centering
    \includegraphics[width=0.9\linewidth]{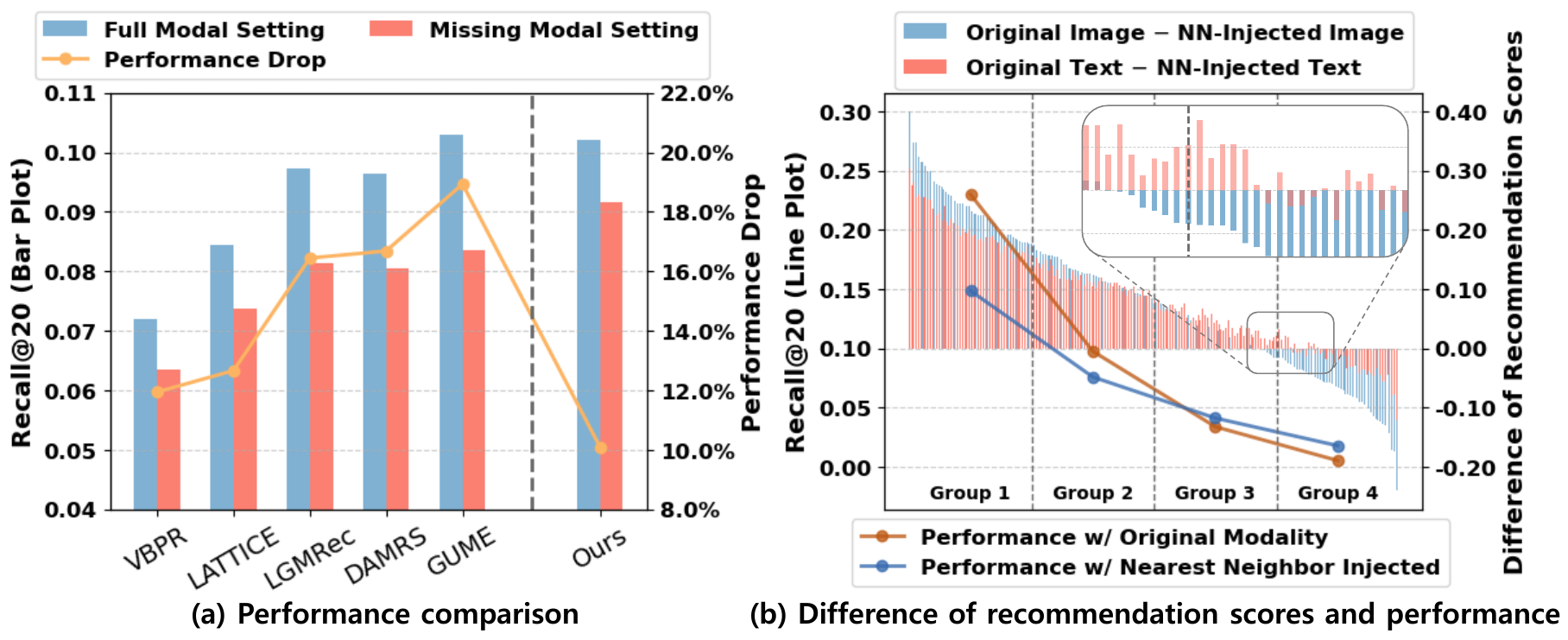} 
    \vspace{-2ex}
    \caption{(a) Performance drop of recent MRSs when missing modality exists. (b) Difference between the model’s recommendation scores under two conditions: when the modality exists (Original Image/Text) and when the modality is missing (NN-Injected Image/Text). Line plots indicate the performance of LGMRec \cite{guo2024lgmrec}.} \label{fig:1_1}
    \vspace{-4ex}
\end{figure}
While collaborative filtering (CF) models have proven successful, they are still constrained by the inherent sparsity of user feedback \cite{zhou2023mmsurvey}. To address this, the integration of diverse modalities, such as images, text, and audio, has emerged as a promising solution, complementing the lack of feedback and enabling a deeper understanding of user preferences compared to traditional CF approaches. This has driven the development of multi-modal recommender systems (MRSs), which leverage multi-modal content to derive rich semantic representations and uncover relationships between items that CF models alone cannot achieve.

Although MRSs have demonstrated their effectiveness, several practical challenges remain insufficiently addressed.

\textbf{C1. Missing modality scenarios are not sufficiently addressed.} Prior MRSs generally assume that all modality features of an item are always fully available. However, in the real-world industry, some or all modality features of an item may be missing \cite{cho2021missing}. 
Specifically, existing studies in MRSs often address the missing modality scenarios by either 1) simply dropping out the items with missing-modality features from the training dataset \cite{zhou2023mmsurvey, wu2024surveyMissing} or 2) injecting synthetic features for missing modalities (e.g., using the global mean of a modality \cite{malitesta2024missingpropar} or the nearest neighbor mean (NN) \cite{malitesta2024dowedrop}). 
Although injection methods are shown to be effective~\cite{malitesta2024dowedrop}, they still encounter a significant performance drop under missing modality scenarios. 
In Figure \ref{fig:1_1}(a), we compared the performance of several MRSs {that assume all modality features are available}, including early models (i.e., VBPR \cite{he2016vbpr} and LATTICE \cite{zhang2021lattice}) as well as state-of-the-art (SOTA) models (i.e., LGMRec \cite{guo2024lgmrec}, DAMRS \cite{xv2024damrs}, and GUME \cite{lin2024gume}), on the Amazon Baby dataset under two scenarios: one where all modalities are available and the other where some modalities are randomly missing\footnote{\label{note1}The missing modality feature is injected based on the NN-injection approach \cite{malitesta2024dowedrop}.}.
{We found that such a naive injection approach still experiences significant performance drop in the presence of missing modalities.}

To investigate why the NN-injection method fails to prevent the performance drop under missing modality, we analyze its impact on the model prediction of a recent MRS, LGMRec \cite{guo2024lgmrec}.
In Figure \ref{fig:1_1}(b), we calculate the difference in recommendation scores of a positive user–item pair between scenarios where the item includes all modality features and where a specific modality feature (either image or text) is missing\textsuperscript{\ref{note1}}, and sort them in descending order.
We also divided the positive user–item pairs into four groups based on their sorted differences, and evaluated the recommendation performance for each group (in the line plot). 
We observe that the drop in the recommendation performance under missing modality (i.e., red line - blue line) is more severe when the difference of the recommendation scores is larger. 
Since a difference of the recommendation scores indicates how well the NN-injected feature captures the original modality (i.e., the smaller the better), 
this result implies that the poor recommendation performance is originated from the NN-injected feature failing to successfully substitute the original modality.
The performance degradation under missing modality exacerbates in recent SOTA MRSs that heavily rely on item modality features. 

\textbf{C2. Unique characteristics of modalities are overlooked.}
Existing MRSs \cite{lin2024gume, guo2024lgmrec} generally aim to directly align among different modalities of an item, assuming that the features of various modalities for the same item inherently share semantics.
However, this assumption does not always hold.
Specifically, the image modality tends to capture visual attributes such as color and style, emphasizing the tangible and aesthetic aspects of an item. In contrast, the text modality conveys descriptive and contextual information, highlighting functional attributes or background details. 
This indicates that each modality contains unique, modality-specific information that cannot be fully captured by other modalities.  For this reason, directly aligning different modalities of an item as in prior studies \cite{lin2024gume, guo2024lgmrec} fails to account for these distinct characteristics, hindering the development of high-quality item representations.

In fact, an existing MRS, LGMRec \cite{guo2024lgmrec}, failing to capture the modality-specific information can be observed by closely examining the difference of the recommendation scores of Group 3 and 4 in Figure \ref{fig:1_1}(b), where the sign of the red and the blue bars are the opposite (See the zoomed part). We argue that the items with the opposite sign in the red and blue bars are those that LGMRec failed to account for the unique characteristics of the text and image modalities. More precisely, as the injected features can be considered as the features commonly shared by all items, the difference in the recommendation scores can be interpreted as the distinct modality-specific information that remains for an item after accounting for the commonly shared features. However, as the injected features cannot be perfectly representative of the globally shared common features of a certain modality, the difference of the recommendation scores would contain not only the {modality-specific information} but also {generally shared information} within a modality of an item.
From this perspective, the positive difference observed for text (red bars) indicates that the remaining information in text is informative to the model's predictions, whereas the negative difference in image (blue bars) suggests the information retained in images less helpful or may even hinder the predictions. This contrast that is present for an item arises because the modality-specific information provided by text and image modalities differs significantly. These observations highlight that each modality contains unique information that is neither shared nor aligned with other modalities. 
Hence, in these cases, forcing alignment between modalities can obscure their unique contributions and adversely affect recommendation performance.

{While there has been some research on missing-modality aware recommender systems (MMA-RSs) \cite{bai2024milk, wang2018lrmm, lin2023ci2mg, ganhor2024sibrar} to address the realistic challenge of missing modalities, these approaches primarily focus on the robustness of the models when handling items with missing modalities. However, because they use the missing modalities in their incomplete state without addressing them, these methods fail to leverage the unique characteristics inherent to each modality. Additionally, since they mainly rely on content-based approaches, they tend to underperform conventional collaborative filtering (CF) methods in general performance, limiting their practical applicability despite their effort to handle real-world challenges.}

{To overcome the inherent limitations of MRSs and MMA-RSs, we propose a novel model called \textbf{D}isentangling and \textbf{G}enerating \textbf{M}odality \textbf{Rec}ommender (\proposed), which effectively addresses the challenges of handling missing modality features and extracting common and unique characteristics of modalities.}

\textbf{C1: To handle missing modality features}, \proposed~ employs an autoencoder architecture to reconstruct an item's modality features to closely resemble the actual ones, ensuring the preservation of the item's distinct attributes. For items with missing modalities, \proposed~ generates modality features by leveraging two sources of knowledge: aligned features from other available modalities and interacted users' modality preferences. Using generated features, \proposed~ enhances the item-item graph, achieving more robust and richer semantic relations between items, which existing models have struggled to capture when missing modality exists.

\textbf{C2: To extract common and unique modality features}, \proposed~ derives general and specific modality features using separate encoders. Additionally, it employs two information-based loss functions to disentangle the general and specific features within a single modality while simultaneously learning shared traits across different modalities.

{These two challenges are closely interconnected, as disentangling modality attributes directly impacts the quality of feature generation for accurately representing an item.}

Our contributions are summarized as follows:
\begin{itemize}[leftmargin=*, itemsep=0pt, topsep=0pt]
    \item We identify and analyze the significant performance degradation of current MRSs in missing modality scenarios, highlighting the inadequacy of naive injection methods as the core limitation.
    \item We propose a robust generation-based approach that reconstructs missing modalities by leveraging an item's distinct characteristics, enabling \proposed~ to achieve superior performance across diverse real-world scenarios.
    \item The proposed fine-grained missing modality feature generation enables \proposed~to perform additional tasks, such as cross-modal retrieval, to assist user behavior in missing scenarios, which remains unattainable for existing recommenders.
    \item We introduce a novel approach from the perspective of mutual information to separate and learn general and specific modality features.
\end{itemize}

\vspace{-2ex}
\section{Related Works}
\subsection{Multi-modal Recommender Systems}
Multi-modal Recommenders (MRSs) leverage multi-modal content in various ways. \textbf{(1) Feature-based} approaches that directly utilize features have been extensively studied. For instance, VBPR \cite{he2016vbpr} integrates visual features directly with ID embeddings. Similarly, methods such as SLMRec \cite{tao2022slmrec}, GRCN \cite{wei2020grcn}, and MMGCN \cite{wei2019mmgcn} use graph convolution networks (GCNs) to integrate modality knowledge with CF knowledge. BM3 \cite{zhou2023bm3} employs contrastive views of modalities for self-supervised learning, while LGMRec \cite{guo2024lgmrec} adopts a hybrid approach by utilizing both hypergraphs and local graphs to balance the learning of global and local knowledge in modality feature extraction.
On the other hand, \textbf{(2) Graph-based} approaches focus on identifying relationships between items based on modality features rather than directly utilizing these features. Notable examples include LATTICE \cite{zhang2021lattice}, MICRO \cite{zhang2022micro}, FREEDOM \cite{zhou2023freedom}, and DAMRS \cite{xv2024damrs}.
Recently, \textbf{(3) Hybrid} approaches combining both strategies have emerged. For instance, GUME \cite{lin2024gume} and MGCN \cite{yu2023mgcn} simultaneously leverage modality features and item-item graphs.

These MRSs have benefited significantly from multi-modal alignment. However, existing methods either completely disregard the relationships between modalities \cite{zhou2023bm3, tao2022slmrec} or rely solely on direct alignment to extract shared information between modalities \cite{lin2024gume, guo2024lgmrec}. As a result, the unique information inherent to different modalities is often overlooked and remains unlearned.
{Some recent works have highlighted the importance of unique modality features for the qualitative representation of items and have adopted orthogonal learning methods from other domains \cite{yu2023mgcn, lin2024gume, yu2021modalspecificMSA}. However, their simplistic approach—taking the mean of modality features as shared commonality and substituting it with the modality features for the unique features—lacks sufficient logical justification from an informational perspective.}
\vspace{-2ex}
\subsection{Missing Modality-Aware Recommender Systems}
Most existing MRSs assume that all modalities are available and complete, which is rarely the case in real-world applications. Missing modality-aware recommender systems (MMA-RSs) address these challenges by considering two common scenarios: (1) \textbf{Incomplete Modality}, where some feature values are missing in a modality \cite{wu2020incomplete2, zhu2021incomplete1, wang2018lrmm}, and (2) \textbf{Missing Modality}, where an entire modality is unavailable \cite{lin2023ci2mg, bai2024milk, ganhor2024sibrar, malitesta2024dowedrop}. MILK \cite{bai2024milk} and SIBRAR \cite{ganhor2024sibrar} tackle missing modalities by leveraging invariant learning and single-branch networks to make robust recommendations without directly addressing the missing modality. Alternatively, \cite{malitesta2024dowedrop} investigates which features can serve as effective substitutes for missing modalities while leaving the model architecture largely unchanged.
{However, most MMA-RS methods not only fail to effectively capture CF knowledge due to their content-based architecture but also rely on naive approaches that lack sufficient consideration of the unique characteristics of each modality, resulting in suboptimal overall performance.}
In this regard, the work most aligned with our motivation is CI2MG \cite{lin2023ci2mg}, which utilizes hypergraphs and optimal transport (OT) to generate missing modalities. However, CI2MG suffers from significant computational overhead in calculating OT and lacks integration between the OT process and other recommendation modules.

\begin{figure*}[t]
    \centering
    \includegraphics[width=0.92\linewidth]{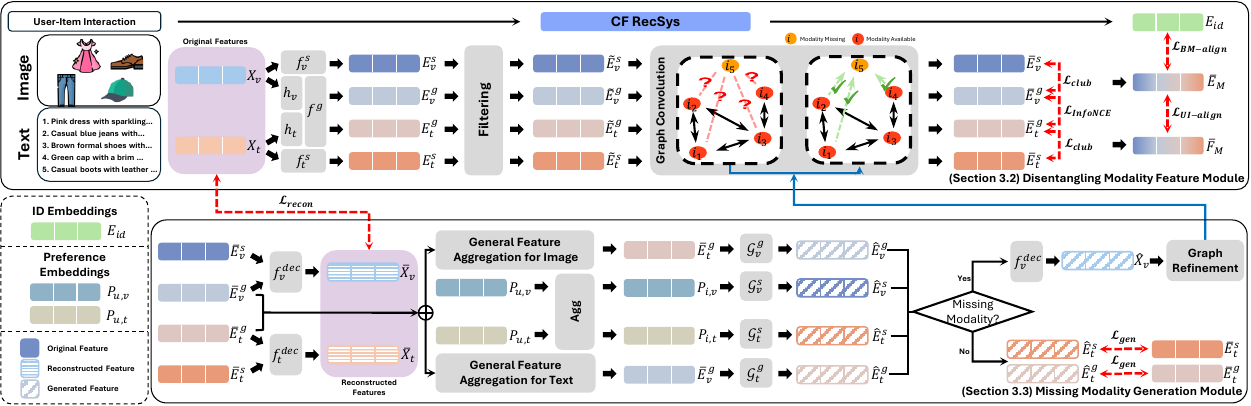} 
    \vspace{-1ex}
    \caption{Overview of \proposed~ framework. It consists of the Disentangling Modality Feature module and the Missing Modality Generation module. In the Missing Modality Generation module, we illustrate the case where an item is associated with the text modality while the image modality is missing.}
    \label{fig:missing-modality}
    \vspace{-2ex}
\end{figure*}

\vspace{-2ex}
\section{Methodology}
To accurately generate missing modalities that reflect the distinct characteristics of items, it is essential to consider both the general (shared) and specific (unique) features of modalities. To achieve this, we propose a \textbf{Disentangling Modality Feature} module (Sec~\ref{method: disentangle}), which uses separate encoders to extract general and specific features. Additionally, we introduce information-based losses to disentangle these features and ensure the alignment of general features across modalities.

Subsequently, to extract meaningful modality representations and construct item-item graphs in the presence of missing modalities, we propose a \textbf{Missing Modality Generation} module (Section~\ref{method: mmgen}), which generates general and specific features that capture the distinct characteristics of items in a fine-grained manner and adaptively refines the item-item graph using these generated features. 

Finally, while modality features are effective in capturing semantic content, they lack the collaborative knowledge critical for recommendation tasks \cite{yuan2023wheretogo, kim2024allmrec}. To mitigate this, we introduce two additional \textbf{alignment methods} (Section~\ref{method: align}) to connect user representations with item representations, thereby bridging collaborative filtering with modality features.

\vspace{-1ex}
\subsection{Preliminaries}
Let \mm{\mc{U}} and \mm{\mc{I}} denote the sets of users and items, with \mm{|\mc{U}|} and \mm{|\mc{I}|} representing the number of users and items, respectively. The User-Item Interaction matrix is defined as \mm{\mc{R} = \mb{R}^{|\mc{U}| \times |\mc{I}|}}. Each item's modality feature is 
{extracted using pre-trained models (i.e., image with a CNN model \cite{he2016cnn} and text with a SBERT \cite{reimers2019sentenceBert})}
and represented as \mm{X_{m} \in \mb{R}^{|\mc{I}| \times d_m}}, where missing modality features are initialized with their mean values. Additionally, the ID embeddings for users and items are defined as \mm{E_{id} \in \mb{R}^{(|\mc{U}| + |\mc{I}|) \times d}}.\footnote{{Matrices are denoted by uppercase letters and vectors by lowercase letters for clarity and consistency (e.g., \mm{e_{i,id}} represents the ID embedding vector for the \mm{i}-th item, and \mm{e_{u,id}} for the \mm{u}-th user, where \mm{i, j} index items and \mm{u, v} index users).}}
{\mm{\mc{N}_i} denotes the set of users that interacted with item \mm{i}, and \mm{\mc{N}_u} denotes the set of items that user \mm{u} interacted with.

\vspace{-1ex}
\subsection{Disentangling Modality Feature Module} \label{method: disentangle}
Existing MRSs directly align modality features \cite{lin2024gume, guo2024lgmrec} or combine adjacency views across modalities \cite{zhang2021lattice, zhou2023freedom}, obscuring unique characteristics of modalities. To address this, we extract two types of features—general and specific—from each modality using encoder functions (Sec~\ref{3.2.1}). These features are further refined through a GCN-based item-item graph and information-driven disentanglement (Sec~\ref{3.2.2}).

\subsubsection{Extracting Modality Features}\label{3.2.1}
To extract general \mm{E^g_{m}} and specific features \mm{E^s_{m}} for each modality $m$, we employ two separate encoders composed of a fully-connected layer: the general encoder \mm{f^g} applicable across all modalities and the specific encoder \mm{f^s_m} tailored to each modality $m$.
\begin{align}
    E^g_m = f^{g}(h_m(X_m)) , \quad E^s_{m} = f^s_{m}(X_{m})
\end{align}
 where the general encoder \mm{f^{g}:\mb{R}^{d} \rightarrow \mb{R}^d} shares parameters across modalities to extract common attributes, while the specific encoder \mm{f^s_m:\mb{R}^{d_m} \rightarrow \mb{R}^d} employs independent parameters to capture the unique characteristics of each modality \mm{m}.
Moreover, since \mm{X_m} has different dimensions across modalities, we use another fully-connected layer \mm{h_m:\mb{R}^{d_m} \rightarrow \mb{R}^d} to project each modality into a unified dimension.

In addition, we introduce a \textbf{Modality Preference Embedding} matrix for users, denoted as \mm{P_{u,m} \in \mb{R}^{|\mc{U}| \times d}}, to more effectively capture users' modality preferences, which is then used to compute the  
item's modality preference matrix \mm{P_{i,m} \in \mb{R}^{|\mc{I}| \times d}} as follows:
\begin{align} \label{filter}
    p_{i,m} = \frac{1}{|\mc{N}_i|}\sum_{u \in \mc{N}_i} p_{u,m}
\end{align} 
where $p_{u,m}\in\mathbb{R}^d$ and $p_{i,m}\in\mathbb{R}^d$ are the modality preference embedding of user $u$ and item $i$, respectively.
As this item preference embedding matrix \mm{P_{i,m}} contains the modality preference of interacted users, we use it to align the modality feature with the user preference as follows:
\footnote{For notational convenience, the superscripts \mm{g} and \mm{s}, denoting general and specific features, are omitted in this equation and subsequent equations where the context makes their meaning clear.}
\begin{align} \label{filter}
    \tilde{E}_{m} = E_{m} \odot \sigma(P_{i,m})
\end{align} 
where $\odot$ is the element-wise product and \mm{\sigma} is the sigmoid function. {We expect the denoised features \mm{\tilde{E}_{m}} to retain information relevant to the user preference, leading to improved recommendations.}

The denoised modality features are then enhanced through GCNs using an item-item graph.
Specifically, we construct an adjacency matrix \mm{S^m} with top-k similar items based on similarity scores of items' modality features \mm{X_{m}} where \mm{S^m} is computed as follows:
\begin{equation} \label{recon_S}
     \begin{aligned}
        S^{m}_{i,j} &= \frac{(x_{i,m})^\top x_{j,m}}{{\|x_{i,m}\| \|x_{j,m}\|}}
    \end{aligned}   
\end{equation}
As the backbone GCNs, we utilize LightGCN \cite{he2020lightgcn} for its computational simplicity and widespread adoption as follows:
\begin{align}
    \bar{E}^{(l)}_{m} = {S}^m \cdot \bar{E}^{(l-1)}_{m}, \text{where}\,\, \bar{E}^{(0)}_{m} = \tilde{E}_{m} \,\text{ and } \,\bar{E}_{m} = \bar{E}^{(L)}_{m}
\end{align}
{where \mm{\bar{E}^{(l)}_{m} \in \mb{R}^{|\mc{I}| \times d}} denotes the modality feature at the \mm{l}-th layer of graph convolution, $L$ is the number of layers. 
Note that we use the last \mm{L}-th layer representation as the item's modality feature.
}
Finally, given the item's final modality feature matrix $\bar{E}^{(L)}_{m}$, we compute the user modality feature matrix \mm{\bar{F}_{m} \in \mb{R}^{|\mc{U}| \times d}} by aggregating the modality features of items the user has interacted with as follows:
\begin{align}
    \bar{f}_{u,m} = \frac{1}{|\mc{N}_u|}\sum_{i \in \mc{N}_u}\bar{e}_{i,m}
\end{align} 

\subsubsection{Disentangling Modality Features}\label{3.2.2}
Separating encoders for general and specific features alone is insufficient to achieve effective disentanglement. Therefore, we introduce information-based approaches leveraging two contrastive losses: one reduces mutual information between general and specific features within a single modality, while the other enhances mutual information between general features across multiple modalities.

\textbf{To minimize the mutual information between general and specific features within the same modality}, we employ a sample-based approach using the Contrastive Log-ratio Upper Bound (CLUB) \cite{cheng2020club} with variational distribution \mm{q_\phi(\cdot|\cdot)} with parameter \mm{\phi} to estimate conditional distribution \mm{p(\cdot|\cdot)}. \mm{q_\phi} consisting of 2-layer MLPs. \begin{align}
    \mc{L}_{club} =   \sum_{i \in \mc{I}} \left[\log q_\phi (\bar{e}^g_{i,m}| \bar{e}^s_{i,m}) - \frac{1}{|\mc{I}|}\sum_{j\in\mc{I}} \log q_\phi (\bar{e}^g_{j,m}| \bar{e}^s_{i,m}) \right]
\end{align}
Following \cite{cheng2020club}, \mm{\mc{L}_{club}} approximates the upper bound of mutual information between \mm{\bar{E}^g_{m}} and \mm{\bar{E}^s_{m}}. We iteratively minimize \mm{\mc{L}_{club}} alongside other model parameters, encouraging modality's general and specific features to have complementary information.

\textbf{To maximize the mutual information between general features of different modalities}, we adopt the InfoNCE loss \cite{oord2018infonce} to approximate a negative lower bound of mutual information. \begin{align} \label{infoNCE}
    \mc{L}_{InfoNCE} = \sum_{i \in \mc{I}} -\log \frac{\exp (\bar{e}^{g}_{i,m} \cdot \bar{e}^g_{i,m'})}{\sum_{j \in \mc{I}} \exp (\bar{e}^{g}_{i, m} \cdot \bar{e}^{g}_{j, m'})}
\end{align}
InfoNCE effectively aligns the general features of different modalities (i.e., \mm{\bar{E}^g_{m}} and \mm{\bar{E}^g_{m'}} for different modalities \mm{m, m'}), by mapping them into the same latent space. Thus, minimizing  InfoNCE loss maximizes the lower bound of mutual information, ensuring better alignment among the general features of multiple modalities.

The final loss for disentanglement is shown below:
\begin{align} \label{eq:disentangle}
    \mc{L}_{disentangle} = \mc{L}_{club} + \mc{L}_{InfoNCE}
\end{align}
{In summary, we expect \mm{\mc{L}_{disentangle}}, to effectively disentangle general and specific features within a modality while aligning general features across different modalities. These losses enable the utilization of cross-modality information to generate missing modality representations, maintaining well-aligned modalities and preserving their unique characteristics}

\subsection{Missing Modality Generation Module} \label{method: mmgen}
As discussed in Section \ref{method: disentangle}, modality features are disentangled into general and specific components using an information-based approach. Building on these well-separated features, we train the decoder to accurately reconstruct raw modality features (Sec \ref{3.3.1}). Then, \proposed~ generates general and specific features for missing modalities in a tailored manner (Sec \ref{3_3_2_MissingModalityGeneration}). These generated features are then utilized to generate raw modality features and refine the item-item graph, addressing the instability caused by missing modalities (Sec \ref{3.3.3}).

\subsubsection{Modality Feature Reconstruction.} \label{3.3.1}
We employ an additional decoder \mm{f^{dec}_m : \mb{R}^d \rightarrow \mb{R}^{d_m}} for each modality $m$ that reconstructs a modality's raw feature \mm{X_{m}} using general features \mm{\bar{E}^g_{m}} and specific features \mm{\bar{E}^s_{m}} as follows:
\begin{align}
    \bar{X}_{m} &= f^{dec}_m (\bar{E}^g_{m} \oplus \bar{E}^s_{m})
\end{align}
where \mm{\oplus} denotes the concatenation operation. The reconstruction process is guided by a reconstruction loss \mm{\mc{L}_{recon}} between raw modality features \mm{X_{m}} and reconstructed modality features \mm{\bar{X}_{m}}.
It ensures that the reconstructed features closely resemble raw feature, allowing \proposed~ to accurately generate features for missing modalities by leveraging the learned \mm{\bar{E}^g_{m}} and \mm{\bar{E}^s_{m}}.
\begin{align} \label{eq:recon}
    \mc{L}_{recon} &= \sum_{i \in \mc{I}} MSE(x_{i,m}, \bar{x}_{i,m}) 
\end{align}
where \mm{MSE} is the mean squared error loss. 
That is, the disentangled features \mm{\bar{E}^g_{m}} and \mm{\bar{E}^s_{m}} retain meaningful modality information to accurately reconstruct \mm{\bar{X}_{m}}, rather than being merely meaningless representations of disentanglement.

\subsubsection{Missing Modality Generation.} \label{3_3_2_MissingModalityGeneration}
To address missing modalities, it is crucial to effectively generate both general and specific features separately, which are used to generate raw modality feature through decoders. 
Although these two features (general and specific) originate from the same modality, they possess fundamentally different characteristics and thus require distinct approaches for generation.

To \textbf{generate general features} \mm{\hat{E}^g_{m}}, we leverage the original general features \mm{\bar{E}^g_{m'}} from other modalities \mm{m'}, aligned by \mm{\mc{L}_{InfoNCE}} in Eq \ref{infoNCE}.
As directly using these features could be unstable, we introduce a general feature generator \mm{\mc{G}^g_m}, which consists of 2-layer MLPs, for each modality $m$ to generate general features as follows:
\begin{align}
    \hat{E}^g_{m} &= \mc{G}^g_{m}(\bigoplus_{m'} \bar{E}^g_{m'})
\end{align}
where \mm{\bigoplus} denotes the concatenation operation applied across all available modalities \mm{m'}.
Note that when the input modality \mm{m'} is also missing (i.e., when two or more modalities are missing), we use the mean of the modality features to ensure that the concatenated vector maintains a consistent dimension regardless of the number of missing modalities.

To \textbf{generate specific features} \mm{\hat{E}^s_{m}}, information from other modalities cannot be utilized.
Instead, we leverage user modality preferences, which are aligned with the items' modality features as shown in Eq \ref{filter}. 
Since the modality-specific knowledge of an item is implicitly captured within the modality preferences of all users associated with that item, a specific feature generator \mm{\mc{G}^s_m} for each modality, which consists of 2-layer MLPs, uses this information to generate the specific features for each item as follows:
\begin{align}
    \hat{E}^s_{m} = \mc{G}^s_{m}(P_{i,m}) 
\end{align}
To guarantee that the generated features preserve the modality's original information, we introduce the generation loss \mm{\mc{L}_{gen}}, which encourages the original features (\mm{\bar{E}^g_{m}}, \mm{\bar{E}^s_{m}}) and the generated features (\mm{\hat{E}^g_{m}}, \mm{\hat{E}^s_{m}}) to be as similar as possible as follows:
\begin{align} \label{eq:gen}
    \mc{L}_{gen} &=  \sum_{i \in \mc{I}} \left( MSE(\bar{e}^{g}_{i,m}, \hat{e}^{g}_{i,m}) + MSE(\bar{e}^{s}_{i,m},  \hat{e}^{s}_{i,m})\right) 
\end{align}
Importantly, these two generation-related losses (\mm{\mc{L}_{recon}} and \mm{\mc{L}_{gen}}) are computed only for items with available modalities, preventing missing modality items from hindering the training process.

\subsubsection{Refining Item-Item Graph via Generated Features.} \label{3.3.3}
At regular intervals determined by a hyperparameter (i.e., every 5 epochs), \proposed~ generates features for items with missing modalities. 
The generated \mm{\hat{E}^g_{m}} and \mm{\hat{E}^s_{m}} are then used to generate \mm{\hat{X}_{m}}, approximating the raw feature \mm{X_{m}} using the decoder as follows:
\begin{align}
    \hat{X}_{m} &= f^{dec}_m (\hat{E}^g_{m} \oplus \hat{E}^s_{m})
\end{align}
However, simply substituting raw modality features with the generated features is insufficient for addressing the issues caused by missing modalities.
As observed in Figure \ref{fig:1_1}, SOTA MRSs that heavily rely on modality features using item-item graphs suffer severe performance degradation.
This occurs because inappropriate edges that are formed when NN features are injected disrupt stable training and hinder the propagation of true semantic relationships.
Hence, we construct a new adjacency matrix \mm{\hat{S}^m} that accurately reflects the semantic relationships among items, based on the generated raw features \mm{\hat{X}_{m}}, following the same process shown in Eq~\ref{recon_S}.

Moreover, to prevent the instability in the model training caused by abrupt graph changes, we propose an adaptive update strategy to smoothly integrate new connections using the adjustable hyperparameter \mm{\alpha} as follows:
\begin{align}
    S^m &= \alpha S^m + (1-\alpha) \hat{S}^m
\end{align}
This is expected to enhance the graph structure and capture richer semantic relationships between items.

{It is important to note that we only update the edges associated with items with missing modalities, and construct directed edges from items containing a modality to those with missing modalities.}

The purpose of this design is twofold:
It prevents contamination of original modality features by under-trained generated features, thereby avoiding instability during training.
Additionally, by computing similarity scores solely for pairs involving items with missing modalities, the process reduces computational overhead, ensuring both stability and efficiency in graph refinement. This approach enables the item-item graph to capture meaningful semantic relationships while maintaining robustness during training.

\smallskip
\noindent \textbf{\underline{Comparisons with prior studies.}}
We would like to emphasize that our approach to refining the item-item Graph differs from the method used in \cite{zhang2021lattice, zhang2022micro}, where the adjacency matrix is first constructed using latent vectors and then jointly optimized with the recommendation loss to enhance performance. Specifically, our approach optimizes a decoder solely to accurately reconstruct raw modality features, which is subsequently used to create the new adjacency matrix.
{When missing modalities exist, our approach can effectively connect edges for items with missing modalities with generated features through the decoder, whereas their method cannot achieve this due to the lack of handling for missing modalities.}

\subsection{Alignment for Recommendation Task} 
\label{method: align}
Modality features effectively capture semantic content but lack the collaborative knowledge essential for recommendation tasks \cite{yuan2023wheretogo, kim2024allmrec}. To address this limitation, we propose two alignment methods that bridge collaborative filtering with modality features. By aligning modality features extracted from pre-trained models with ID embeddings, we seamlessly integrate modality knowledge and collaborative knowledge.

\subsubsection{Fusing ID embedding and Modality Features}
The final modality representation is computed by first obtaining the general features of all modalities through mean pooling. Subsequently, these combined general features are mean pooled with the specific features of the modalities to produce the final modality representation.
\begin{equation}
    \begin{aligned} 
    &\bar{E}_{M} = MeanPool_{m \in \mc{M}}(\bar{E}^s_{m}, MeanPool_{m' \in \mc{M}}(\bar{E}^g_{m'})) \\
    &\bar{F}_{M} = MeanPool_{m \in \mc{M}}(\bar{F}^s_{m}, MeanPool_{m' \in \mc{M}}(\bar{F}^g_{m'}))
    \end{aligned}
\end{equation}
{where $\mathcal{M}$ is a set of all modalities.}
Incorporating modality features, the final recommendation representation for the user and item is derived by combining the ID embedding from the CF model with the respective modality representations as follows:
\begin{equation}
    \begin{aligned}
    \bar{E}_{u} = E_{u, id} + \bar{F}_{M}, \quad \bar{E}_{i} = E_{i, id} + \bar{E}_{M}
    \end{aligned}
\end{equation}
The final recommendation score \mm{y_{u,i}} is computed by the inner product between the user and item representations as follows:
\begin{align}
    y_{u,i} = \bar{e}^\top_u \bar{e}_i
\end{align}

\subsubsection{Behavior-Modality Alignment}
While modality features are important for item representations, the lack of collaborative knowledge makes it difficult to capture user-item relationships \cite{yuan2023wheretogo, kim2024allmrec}.
To alleviate this challenge, we incorporate collaborative knowledge by aligning user behavior with modality features through a contrastive loss.
\begin{equation}
    \begin{aligned}
        \mc{L}_{BM\text{-}align} = 
        \sum_{u \in \mc{U}} -&\log \frac{\exp (e_{u, id} \cdot \bar{f}_{u, M})}{\sum_{v \in \mc{U}} \exp (e_{u, id} \cdot \bar{f}_{v, M})} \\ 
        + & \sum_{i \in \mc{I}} -\log \frac{\exp (e_{i, id} \cdot \bar{e}_{i, M})}{\sum_{j \in \mc{I}} \exp (e_{i, id} \cdot \bar{e}_{j, M})}
    \end{aligned}    
\end{equation}

\subsubsection{User-Item Alignment}
We further refine modality features among users and items to ensure coherence within each modality as follows:
\begin{align}
    \mc{L}_{UI\text{-}align} &= \sum_{m\in \mc{M}}\sum_{(u,i) \in \mc{O}} -\log \frac{\exp (\bar{f}_{u, m} \cdot \bar{e}_{i, m})}{\sum_{j \in \mc{I}} \exp (\bar{f}_{u, m} \cdot \bar{e}_{j, m})}
\end{align}
{where \mm{\mc{O}} is a set of positive observation in interaction matrix \mm{\mc{R}} (i.e., $(u,i)$ is contained in \mm{\mc{O}} when \mm{\mc{R}_{u,i} = 1})}.

By aligning the modality features of an item with those of users who interacted with it, \proposed~ captures recommendation-relevant knowledge, enhancing the consistency of the modality features.

The final alignment loss is defined as:
\begin{align} \label{eq:align}
    \mc{L}_{align} &= \mc{L}_{BM\text{-}align} + \mc{L}_{UI\text{-}align}
\end{align}
By combining \mm{\mc{L}_{BM\text{-}align}} and \mm{\mc{L}_{UI\text{-}align}}, our approach effectively integrates modality features with collaborative filtering knowledge, resulting in a robust and comprehensive recommendation system.

To optimize user and item representations for the recommendation task, we employ the Bayesian Personalized Ranking (BPR) loss \cite{rendle2012bpr}, 
\begin{align}
    \mc{L}_{bpr} = \sum_{(u,i^+, i^-) \in \mc{D}} \left( -\sigma(y_{u,i^+} - y_{u,i^-}) \right)
\end{align}
{where \mm{\mc{D} = \{(u, i^+, i^-)| (u, i^+) \in \mc{O}, (u, i^-) \notin \mc{O}\}} represents the dataset of triplets, and \mm{\sigma()} denotes the sigmoid function.}

The final objective function of~\proposed~is give by:
\begin{align}
    \mc{L} = \mc{L}_{bpr} + \mc{L}_{recon} &+ \mc{L}_{gen} + \lambda_1 \mc{L}_{disentangle} + \lambda_2 \mc{L}_{align}
\end{align}
where \mm{\lambda_1, \lambda_2} are hyper-parameters.
\renewcommand{\arraystretch}{0.9}
\begin{table}[t!]
\centering
\resizebox{0.92\linewidth}{!}{%
\begin{tabular}{l|cccc|ccc}
\toprule
\multirow{2}{*}{Dataset} & \multirow{2}{*}{\# Users} & \multirow{2}{*}{\# Items} & \multirow{2}{*}{\# Interactions} & \multirow{2}{*}{Sparsity} & \multicolumn{3}{c}{Modalities} \\ 
                         &                           &                           &                                  &                           & Image     & Text    & Audio    \\ \midrule
Baby                     & 19,445                    & 7,050                     & 160,792                          & 99.88\%                   & \cmark    & \cmark  & \xmark   \\
Sports                   & 35,598                    & 18,357                    & 296,337                          & 99.95\%                   & \cmark    & \cmark  & \xmark   \\
Clothing                 & 39,387                    & 23,033                    & 278,677                          & 99.97\%                   & \cmark    & \cmark  & \xmark   \\
TikTok                   & 9,308                     & 6,710                     & 68,722                           & 99.89\%                   & \cmark    & \cmark  & \cmark   \\
\bottomrule
\end{tabular}}
\caption{Statistics of Datasets} \label{tab:dataset_stats}
\vspace{-7ex}
\end{table}
\begin{table*}[t!]
\centering
\caption{Performance Comparison. The best and runner-ups are marked in bold and underlined, respectively.}
\label{tab:performance}
\vspace{-2ex}
\resizebox{0.91\linewidth}{!}{
\begin{tabular}{cc|cccc|cccc|cccc|cccc} \toprule
\multicolumn{18}{c}{\textbf{Missing Modality Setting}} \\ \midrule
\multicolumn{2}{c|}{Dataset} & \multicolumn{4}{c|}{Baby} & \multicolumn{4}{c|}{Sports} & \multicolumn{4}{c|}{Clothing} & \multicolumn{4}{c}{TikTok} \\ \midrule
\multicolumn{2}{c|}{Metric} & R@20 & R@50 & N@20 & N@50 & R@20 & R@50 & N@20 & N@50 & R@20 & R@50 & N@20 & N@50 & R@20 & R@50 & N@20 & N@50 \\ \midrule 
\multirow{5}{*}{\rotatebox{90}{CF}}
&   MF        & 0.0611 & 0.1091 & 0.0273 & 0.0370 & 0.0707 & 0.1112 & 0.0327 & 0.0416 & 0.0346 & 0.0533 & 0.0164 & 0.0201 & 0.0558 & 0.0909 & 0.0220 & 0.0289 \\
&   NGCF      & 0.0602 & 0.1137 & 0.0258 & 0.0366 & 0.0701 & 0.1215 & 0.0304 & 0.0408 & 0.0422 & 0.0728 & 0.0176 & 0.0237 & 0.0722 & 0.1284 & 0.0284 & 0.0394 \\
&   LightGCN  & 0.0733 & 0.1323 & 0.0320 & 0.0440 & 0.0829 & 0.1369 & 0.0379 & 0.0488 & 0.0514 & 0.0818 & 0.0227 & 0.0288 & 0.0916 & 0.1576 & 0.0406 & 0.0536 \\
&   SGL       & 0.0804 & 0.1422 & 0.0348 & 0.0473 & 0.0917 & 0.1492 & 0.0414 & 0.0531 & 0.0600 & 0.0936 & 0.0271 & 0.0338 & 0.0939 & 0.1490 & 0.0403 & 0.0521 \\
&   SimGCL    & 0.0809 & 0.1409 & 0.0349 & 0.0471 & 0.0910 & 0.1465 & 0.0410 & 0.0523 & 0.0542 & 0.0833 & 0.0252 & 0.0310 & 0.0952 & 0.1451 & 0.0401 & 0.0509 \\ \midrule
\multirow{10}{*}{\rotatebox[origin=c]{90}{        
            \begin{tabular}{c}
                Multi-Modal \\
                Recommenders
            \end{tabular}
                }}
&   VBPR      & 0.0514 & 0.0937 & 0.0213 & 0.0299 & 0.0741 & 0.1229 & 0.0328 & 0.0427 & 0.0462 & 0.0737 & 0.0207 & 0.0226 & 0.0410 & 0.0699 & 0.0172 & 0.0229 \\
&   MMGCN     & 0.0519 & 0.0991 & 0.0215 & 0.0310 & 0.0509 & 0.0913 & 0.0215 & 0.0297 & 0.0289 & 0.0530 & 0.0120 & 0.0168 & 0.0883 & 0.1431 & 0.0372 & 0.0484 \\
&   GRCN      & 0.0644 & 0.1151 & 0.0274 & 0.0377 & 0.0681 & 0.1157 & 0.0300 & 0.0397 & 0.0381 & 0.0644 & 0.0161 & 0.0214 & 0.0716 & 0.1257 & 0.0283 & 0.0389 \\
&   SLMRec    & 0.0753 & 0.1254 & 0.0340 & 0.0422 & 0.0914 & 0.1462 & 0.0415 & 0.0526 & 0.0624 & 0.0979 & 0.0281 & 0.0351 & 0.0932 & 0.1523 & 0.0364 & 0.0480 \\
&   BM3       & 0.0683 & 0.1235 & 0.0296 & 0.0408 & 0.0908 & 0.1466 & 0.0400 & 0.0513 & 0.0591 & 0.0920 & 0.0268 & 0.0334 & 0.0768 & 0.1215 & 0.0322 & 0.0409 \\
&   LATTICE   & 0.0738 & 0.1297 & 0.0319 & 0.0432 & 0.0867 & 0.1401 & 0.0306 & 0.0384 & 0.0581 & 0.0929 & 0.0262 & 0.0332 & 0.0824 & 0.1353 & 0.0372 & 0.0477 \\
&   MGCN      & 0.0833 & 0.1389 & 0.0366 & 0.0481 & 0.0941 & 0.1525 & \ul{0.0425} & 0.0544 & 0.0665 & 0.1052 & 0.0300 & 0.0377 & 0.0870 & 0.1395 & 0.0356 & 0.0460 \\
&   LGMRec    & 0.0813 & 0.1410 & 0.0352 & 0.0471 & 0.0906 & 0.1496 & 0.0403 & 0.0522 & 0.0624 & 0.1015 & 0.0277 & 0.0355 & 0.0791 & 0.1376 & 0.0335 & 0.0450 \\
&   DAMRS     & 0.0804 & 0.1390 & 0.0355 & 0.0474 & 0.0941 & 0.1526 & 0.0416 & 0.0534 & \ul{0.0670} & \ul{0.1066} & \ul{0.0301} & \ul{0.0380} & \ul{0.1044} & 0.1638 & \ul{0.0452} & \ul{0.0569} \\
&   GUME      & \ul{0.0835} & \ul{0.1429} & \ul{0.0369} & \ul{0.0489} & \ul{0.0947} & \ul{0.1554} & 0.0424 & \ul{0.0546} & 0.0639 & 0.1016 & 0.0291 & 0.0366 & 0.0968 & \ul{0.1645} & 0.0389 & 0.0524 \\ \midrule
\multirow{3}{*}{\rotatebox[origin=c]{90}{        
            \begin{tabular}{@{}c@{}}
                MMA \\
                RSs
            \end{tabular}
                }}
&   CI2MG     & 0.0720 & 0.1285 & 0.0305 & 0.0420 & 0.0717 & 0.1179 & 0.0331 & 0.0425 & 0.0523 & 0.0845 & 0.0237 & 0.0301 & 0.0772 & 0.1284 & 0.0327 & 0.0429 \\
&   MILK      & 0.0427 & 0.0763 & 0.0182 & 0.0250 & 0.0362 & 0.0626 & 0.0155 & 0.0209 & 0.0226 & 0.0376 & 0.0094 & 0.0124 & 0.0404 & 0.0640 & 0.0184 & 0.0230 \\ 
&   SIBRAR    & 0.0480 & 0.0888 & 0.0207 & 0.0289 & 0.0434 & 0.0758 & 0.0190 & 0.0255 & 0.0264 & 0.0453 & 0.0110 & 0.0148 & 0.0548 & 0.0854 & 0.0220 & 0.0280 \\ \midrule
&   \proposed    & \textbf{0.0897} & \textbf{0.1531} & \textbf{0.0404} & \textbf{0.0528} & \textbf{0.1024} & \textbf{0.1625} & \textbf{0.0462} & \textbf{0.0584} & \textbf{0.0725} & \textbf{0.1134} & \textbf{0.0324} & \textbf{0.0406} & \textbf{0.1093} & \textbf{0.1773} & \textbf{0.0476} & \textbf{0.0611} \\
&   Improv.   & 7.43\% & 7.14\% & 9.49\% & 7.98\% & 8.13\% & 4.57\% & 8.71\% & 6.96\% & 8.21\% & 6.00\% & 7.64\% & 6.84\% & 4.69\% & 7.78\% & 5.31\% & 7.38\%  \\
\bottomrule
\toprule
\multicolumn{18}{c}{\textbf{Missing Modality + New Items Setting}} \\ \midrule
\multicolumn{2}{c|}{Dataset} & \multicolumn{4}{c|}{Baby} & \multicolumn{4}{c|}{Sports} & \multicolumn{4}{c|}{Clothing} & \multicolumn{4}{c}{TikTok} \\ \midrule
\multicolumn{2}{c|}{Metric} & R@20 & R@50 & N@20 & N@50 & R@20 & R@50 & N@20 & N@50 & R@20 & R@50 & N@20 & N@50 & R@20 & R@50 & N@20 & N@50 \\ \midrule 
\multirow{5}{*}{\rotatebox{90}{CF}} 
& MF         & 0.0349 & 0.0583 & 0.0174 & 0.0228 & 0.0376 & 0.0598 & 0.0201 & 0.0253 & 0.0196 & 0.0288 & 0.0100 & 0.0120 & 0.0286 & 0.0464 & 0.0112 & 0.0148 \\
& NGCF       & 0.0336 & 0.0599 & 0.0160 & 0.0220 & 0.0389 & 0.0648 & 0.0196 & 0.0255 & 0.0262 & 0.0418 & 0.0124 & 0.0158 & 0.0443 & 0.0737 & 0.0186 & 0.0246 \\
& LightGCN   & 0.0434 & 0.0723 & 0.0218 & 0.0285 & 0.0458 & 0.0788 & 0.0240 & 0.0302 & 0.0290 & 0.0457 & 0.0147 & 0.0184 & 0.0527 & 0.0829 & 0.0245 & 0.0306 \\
& SGL        & 0.0434 & 0.0682 & 0.0228 & 0.0285 & 0.0484 & 0.0788 & 0.0251 & 0.0316 & 0.0337 & 0.0505 & 0.0173 & 0.0210 & 0.0548 & 0.0775 & 0.0247 & 0.0293 \\
& SimGCL     & 0.0391 & 0.0630 & 0.0208 & 0.0262 & 0.0475 & 0.0750 & 0.0242 & 0.0313 & 0.0307 & 0.0492 & 0.0159 & 0.0200 & 0.0550 & 0.0771 & 0.0245 & 0.0289 \\ \midrule
\multirow{10}{*}{\rotatebox{90}{        
            \begin{tabular}{c}
                Multi-Modal \\
                Recommenders
            \end{tabular}
                }}
& VBPR       & 0.0347 & 0.0640 & 0.0177 & 0.0244 & 0.0393 & 0.0641 & 0.0200 & 0.0257 & 0.0265 & 0.0414 & 0.0133 & 0.0166 & 0.0244 & 0.0417 & 0.0118 & 0.0221 \\
& MMGCN      & 0.0326 & 0.0596 & 0.0157 & 0.0218 & 0.0274 & 0.0489 & 0.0133 & 0.0182 & 0.0170 & 0.0308 & 0.0079 & 0.0110 & 0.0439 & 0.0599 & 0.0186 & 0.0218 \\
& GRCN       & 0.0347 & 0.0621 & 0.0170 & 0.0233 & 0.0368 & 0.0606 & 0.0185 & 0.0239 & 0.0226 & 0.0379 & 0.0109 & 0.0143 & 0.0378 & 0.0661 & 0.0166 & 0.0223 \\
& SLMRec    & 0.0434 & 0.0702 & 0.0223 & 0.0284 & 0.0477 & 0.0755 & 0.0245 & 0.0308 & 0.0344 & 0.0526 & 0.0176 & 0.0217 & 0.0548 & 0.0775 & 0.0247 & 0.0293 \\
& BM3        & 0.0407 & 0.0717 & 0.0204 & 0.0274 & \ul{0.0496} & \ul{0.0796} & \ul{0.0255} & \ul{0.0324} & 0.0317 & 0.0496 & 0.0163 & 0.0202 & 0.0588 & 0.0869 & 0.0262 & 0.0318 \\
& LATTICE    & 0.0423 & 0.0730 & 0.0213 & 0.0283 & 0.0432 & 0.0713 & 0.0218 & 0.0283 & 0.0344 & 0.0539 & 0.0173 & 0.0216 & 0.0444 & 0.0742 & 0.0209 & 0.0269 \\ 
& MGCN       & 0.0446 & \ul{0.0802} & \ul{0.0230} & 0.0302 & 0.0478 & 0.0775 & 0.0240 & 0.0316 & 0.0358 & 0.0562 & 0.0182 & 0.0228 & 0.0357 & 0.0694 & 0.0140 & 0.0208 \\ 
& LGMRec     & 0.0450 & 0.0772 & \ul{0.0230} & 0.0303 & 0.0462 & 0.0742 & 0.0236 & 0.0300 & 0.0353 & 0.0557 & 0.0175 & 0.0221 & 0.0388 & 0.0632 & 0.0142 & 0.0191 \\ 
& DAMRS      & \ul{0.0455} & 0.0779 & 0.0229 & \ul{0.0304} & 0.0480 & 0.0784 & 0.0248 & 0.0317 & \ul{0.0380} & \ul{0.0583} & \ul{0.0192} & \ul{0.0237} & \ul{0.0598} & 0.0872 & \ul{0.0267} & \ul{0.0333} \\ 
& GUME       & 0.0447 & 0.0795 & 0.0225 & \ul{0.0304} & 0.0476 & 0.0776 & 0.0244 & 0.0313 & 0.0357 & 0.0563 & 0.0179 & 0.0224 & 0.0567 & \ul{0.0918} & 0.0217 & 0.0289 \\ \midrule
\multirow{3}{*}{\rotatebox[origin=c]{90}{        
            \begin{tabular}{@{}c@{}}
                MMA \\
                RSs
            \end{tabular}
                }}
& CI2MG      & 0.0415 & 0.0716 & 0.0210 & 0.0279 & 0.0437 & 0.0718 & 0.0226 & 0.0290 & 0.0294 & 0.0461 & 0.0149 & 0.0186 & 0.0427 & 0.0660 & 0.0188 & 0.0235 \\ 
& MILK       & 0.0247 & 0.0429 & 0.0120 & 0.0162 & 0.0192 & 0.0323 & 0.0093 & 0.0123 & 0.0133 & 0.0226 & 0.0064 & 0.0085 & 0.0212 & 0.0332 & 0.0105 & 0.0129 \\ 
& SIBRAR     & 0.0280 & 0.0495 & 0.0138 & 0.0188 & 0.0257 & 0.0435 & 0.0128 & 0.0169 & 0.0153 & 0.0259 & 0.0070 & 0.0094 & 0.0351 & 0.0527 & 0.0154 & 0.0190 \\ \midrule
& \proposed     & \textbf{0.0519} & \textbf{0.0876} & \textbf{0.0257} & \textbf{0.0336} & \textbf{0.0532} & \textbf{0.0845} & \textbf{0.0276} & \textbf{0.0348} & \textbf{0.0413} & \textbf{0.0631} & \textbf{0.0211} & \textbf{0.0260} & \textbf{0.0639} & \textbf{0.0973} & \textbf{0.0285} & \textbf{0.0353} \\
& Improv.    & 14.06\% & 9.22\% & 11.73\% & 10.53\% & 7.26\% & 6.16\% & 8.67\% & 7.41\% & 8.68\% & 8.23\% & 9.89\% & 9.70\% & 6.86\% & 5.99\% & 6.74\% & 6.00\% \\
\bottomrule
\end{tabular}}
\vspace{-2ex}
\end{table*}

\section{Experiment}
\subsection{Experimental Settings}

\textbf{Datasets.} We use datasets with diverse modalities, including the Amazon Baby, Sports, and Clothing datasets, as well as the TikTok dataset with 5-core setting following previous works \cite{wei2023mmssl, zhang2021lattice, xv2024damrs, lin2024gume}.
{The Amazon datasets contain image and text modalities. Modality features in Amazon datasets are extracted using the same pre-trained models following \cite{zhou2023mmrec} (i.e., image with a CNN model \cite{he2016cnn} and text with a SBERT \cite{reimers2019sentenceBert}).
The TikTok dataset, published by TikTok\footnote{\href{https://www.tiktok.com/}{https://www.tiktok.com/}}, includes image, text, and audio modalities. However, the raw features and pre-trained models used for feature extraction are not publicly available.}
The statistics of datasets are summarized in Table \ref{tab:dataset_stats}. 

\smallskip
\noindent\textbf{Missing Modality Setting.} Similar to other MMA-RSs \cite{lin2023ci2mg, bai2024milk, ganhor2024sibrar}, we introduce settings where missing modalities are present. For datasets with two modalities, items were evenly divided such that 1/3 of items have 0, 1, or 2 missing modalities. For three modalities, 1/4 of items have 0, 1, 2, or 3 missing modalities. The specific modality chosen as missing was randomly selected for each item.

\smallskip
\noindent\textbf{New Items Setting.} Following the setup of \cite{bai2024milk}, we select 20\% of the items as new items that appeared only in the test set, ensuring these items were unseen during the training and validation phases. This setup evaluates the model's ability to generalize to previously unseen items{, testing its performance in realistic scenarios.}

\smallskip
\noindent\textbf{Compared Methods.} To ensure fair comparisons, we evaluated \proposed~against a wide range of models, including 5 traditional CF models, 10 multi-modal RSs, and 3 missing modality-aware RSs.
\begin{itemize}[leftmargin=*, itemsep=0pt, topsep=0pt]
    \item \textbf{Traditional CF Models}: Matrix factorization methods (MFBPR \cite{rendle2012bpr}), GNN-based methods (NGCF \cite{wang2019ngcf} and LightGCN \cite{he2020lightgcn}), and contrastive learning methods (SGL \cite{wu2021sgl} and SimGCL \cite{yu2022simgcl}).
    \item \textbf{Multi-modal Recommender Systems}\footnote{The missing modality feature is injected based on the NN-injection approach \cite{malitesta2024dowedrop}.}: Feature-based methods (VBPR \cite{he2016vbpr}, MMGCN \cite{wei2019mmgcn}, GRCN \cite{wei2020grcn}, SLMRec \cite{tao2022slmrec}, BM3 \cite{zhou2023bm3} and LGMRec \cite{guo2024lgmrec}), Graph-based methods (LATTICE \cite{zhang2021lattice} and DAMRS \cite{xv2024damrs}) and Hybrid methods (MGCN \cite{yu2023mgcn} and GUME \cite{lin2024gume}).
    \item \textbf{Missing Modality-Aware Recommender Systems}: Robust learning methods that utilize invariant learning (MILK \cite{bai2024milk}) and single-branch network (SIBRAR \cite{ganhor2024sibrar}), and generation-based methods that utilize Optimal Transport and hypergraphs (CI2MG \cite{lin2023ci2mg}).
\end{itemize}

\smallskip
\noindent\textbf{Evaluation Metrics.} We split each dataset into training, validation, and test sets with 8:1:1 following \cite{xv2024damrs, lin2024gume, guo2024lgmrec}. For evaluation, we employ the widely used Recall@K and NDCG@K metrics with K = 20 and 50. We denote Recall and NDCG as R and N, respectively.

\smallskip
\noindent\textbf{Implementation}
We implement \proposed~ and other baselines in PyTorch \cite{paszke2019pytorch}. We adopt Adam \cite{diederik2014adam} as optimizer. The embedding dimension is fixed to 64. For~\proposed, the hyperparameters \mm{\lambda_1} and \mm{\lambda_2} are tuned in \{1.0, 1e-1, 1e-2, 1e-3\}, and \mm{\tau} in contrastive learning and \mm{\alpha} for adjusting graph structure are tuned from [0,1] {with an interval of 0.2}. The interval of modality generation is fixed to 5 epochs. For convergence consideration, the early stopping and total epochs are fixed at 30 and 1,000, respectively.

\subsection{Performance Comparison} 
For comprehensive evaluations of \proposed, we perform evaluations under various scenarios: overall performance (Sec \ref{sec4_2_1}), different missing modality levels (Sec \ref{sec4_2_2}), varying missing ratios (Sec \ref{sec4_2_3}), and cross-modal retrieval (Sec \ref{sec4_2_4}).

\subsubsection{Overall Performance} \label{sec4_2_1}
In Table \ref{tab:performance}, we present the performance of \proposed~ and other models under the \textbf{Missing Modality Setting}. Additionally, to simulate a more realistic and challenging scenario, we evaluate performance under the \textbf{Missing Modality and New Items Setting}, which combines missing modalities with new items.
\begin{figure}[t!]
    \centering 
    \includegraphics[width=0.9\linewidth]{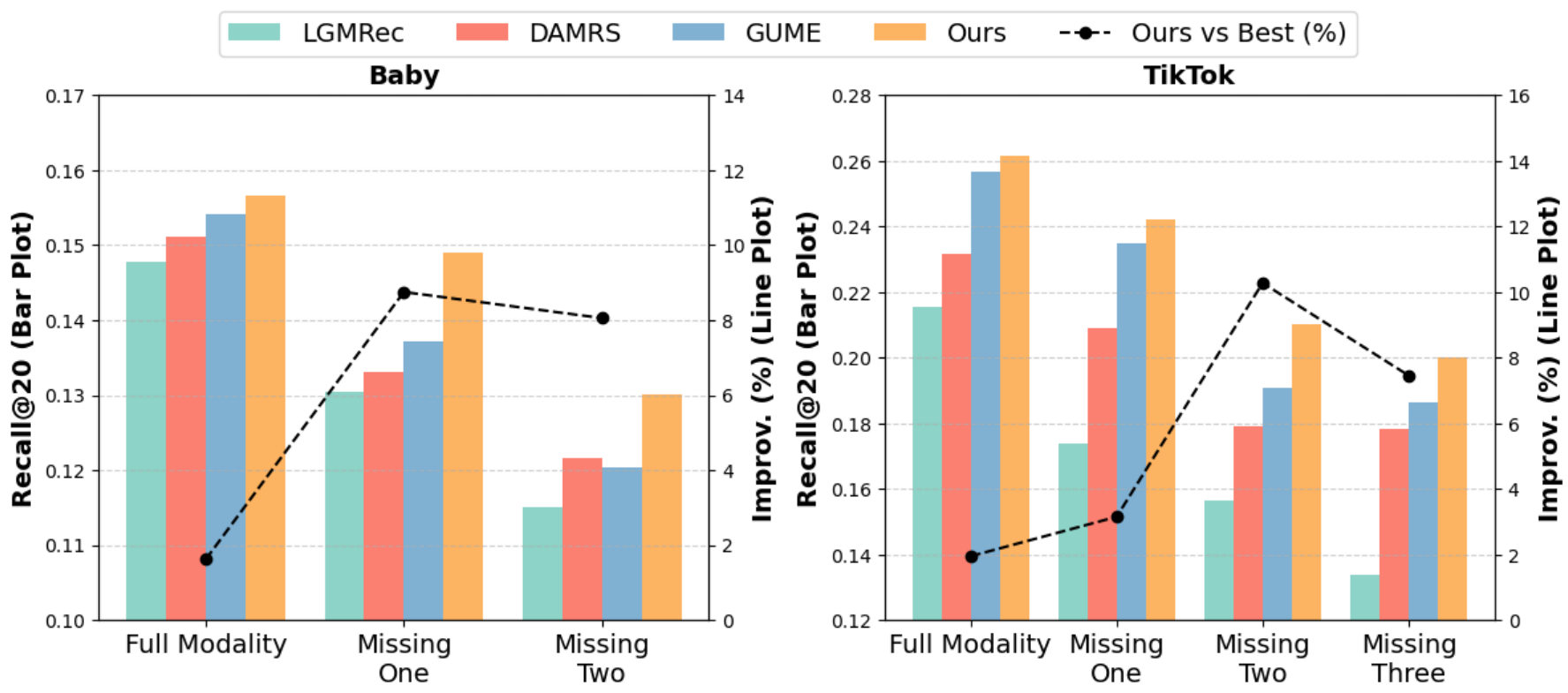} 
    \vspace{-3mm}
    \caption{Performance on various missing levels on Amazon Baby and TikTok datasets.} 
    \label{fig:4_3_1} 
    \vspace{-5mm}
\end{figure}
We have the following observations:
\textbf{1)} \proposed~ consistently outperforms existing MRSs and MMA-RSs across all datasets in both settings.
This highlights the benefit of the proposed generation-based approach to handle missing modalities compared with existing injection-based methods.
By dynamically generating missing modality features, \proposed~ achieves robust and superior performance compared to baseline methods.
\textbf{2)} {The performance gap between the best CF model and the best MRSs narrows when comparing the Missing Modality Setting with the combined Missing Modality and New Item Setting, while the improvement achieved by \proposed~increases. This demonstrates the effectiveness of \proposed's modality generation module in handling new items.}
\textbf{3)} {The MMA-RSs based on contents, MILK and SIBRAR, exhibit lower performance compared to CF models. Additionally, CI2MG, which leverages CF knowledge by utilizing LightGCN as its backbone, performs worse than the vanilla LightGCN. This indicates that generating modality features without proper alignment (i.e., Optimal Transport) can lead to performance degradation.}
\subsubsection{Performance at Different Missing Modality Levels} \label{sec4_2_2}
In Figure \ref{fig:4_3_1}, we evaluate the performance of \proposed~across different levels of missing modalities.
The missing modality settings were consistent with those used in Section \ref{sec4_2_1}, and comparisons were made with LGMRec, DAMRS, and GUME.
For better analysis, we reported the performance improvement of \proposed~relative to the best-performing baseline in a line plot.
\textbf{1)} \proposed~consistently outperformed all baselines across all levels of missing modalities.
This demonstrates that \proposed~ not only obtains meaningful representations through disentangling modality features but also effectively generates missing modalities to enhance performance.
\textbf{2)} Performance gains were greater when some modalities were available compared to cases with no modalities {(i.e., missing one in Baby dataset and missing one \& two in TikTok dataset)}.
This highlights the importance of utilizing other available modalities to generate general features during the generation process.
\textbf{3)} Even in scenarios where no modalities were available {(i.e., missing two in Baby and missing three in TikTok)}, \proposed~ still achieved significant performance improvements. This demonstrates that even in situations where general features cannot be generated due to the complete absence of available modalities, the specific features generated using the preference-based approach with interaction data remain effective.

\subsubsection{Performance under Varying Missing Ratios} \label{sec4_2_3}
We conduct experiments with missing ratios set to 0\% (No Missing Modality), 20\%, 40\%, 60\%, and up to an extreme scenario of 80\%. For each setting, items with missing modalities and the types of missing modalities were randomly selected, with selections kept consistent across experiments.
{For example, if an item's image modality was missing at 20\%, the same item's image modality remained missing at 40\%.}
\begin{figure}[t!]
    \centering
    \includegraphics[width=0.9\linewidth]{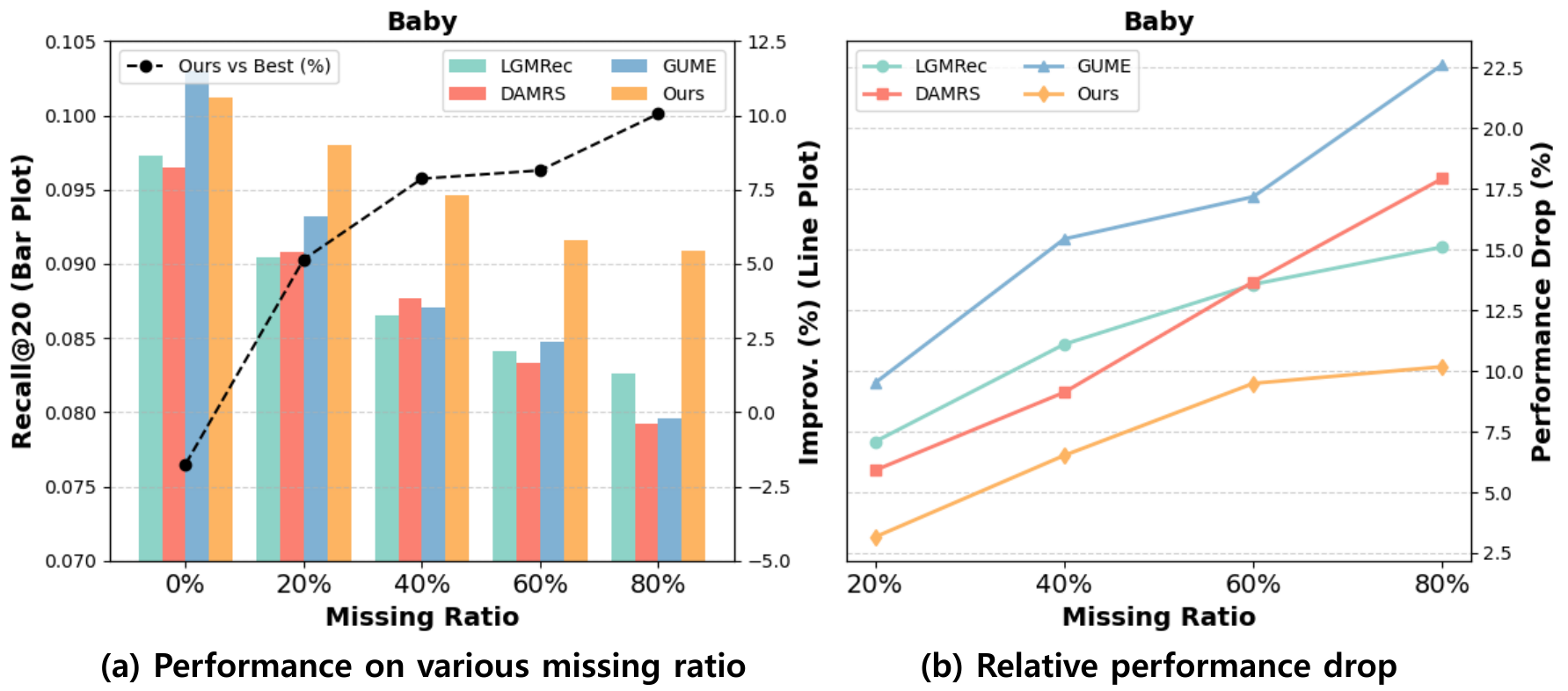} 
    \vspace{-4mm}
    \caption{{(a) Performance on various missing ratios, and (b) relative performance drop on Amazon Baby dataset.}
    } 
    \label{fig:4_3_2} 
    \vspace{-4mm}
\end{figure}
In Figure \ref{fig:4_3_2}(a), we present the results of \proposed~ in comparison with LGMRec, DAMRS, and GUME on the Amazon Baby dataset using a bar plot. Additionally, the performance improvement of \proposed~ relative to the best-performing baseline is shown in a line plot.
\textbf{1)} Across all missing ratios, \proposed~ consistently outperformed all baselines, except for the 0\% baseline{, where its performance was only slightly lower}. Notably, as GUME's performance dropped significantly with increasing missing ratios, the performance gap between \proposed~ and the other models widened substantially, reaching up to 10.05\% at a missing ratio of 80\%. This demonstrates \proposed's robustness across a wide range of missing ratios, from low levels to extreme cases.
To further confirm that \proposed's robust performance is not solely attributed to its high performance at the 0\% missing ratio (No Missing Modality), we also show the percentage performance drop at each missing ratio relative to the 0\% baseline in figure \ref{fig:4_3_2}(b).
\textbf{2)} The performance drop of \proposed~ was consistently smaller compared to other models {, with 2.7\% (3.2\% vs. 5.9\%) at 20\% missing and 4.9\% (10.2\% vs. 15.1\%) at a missing ratio of 80\%}.
{This highlights that \proposed~ not only achieves high performance by effectively leveraging modality features through the {Disentangling Modality Feature} module but also demonstrates robust capability in addressing missing modalities via the {Missing Modality Generation} module.}

\begin{table}[h]
    \centering
    \vspace{-2mm}
    \caption{Results (Hit@10 and Hit@20) on Cross-Modal Retrieval of \proposed. NN means the Nearest Neighbor method.}
    \vspace{-2mm}
    \resizebox{0.9 \columnwidth}{!}{%
    \begin{tabular}{c|cc|cc}
    \toprule
    Datasets & \multicolumn{4}{c}{Baby} \\ \midrule
    &  \multicolumn{2}{c|}{Missing 1 Modality} & \multicolumn{2}{c}{Missing 2 Modalities} \\ \midrule
    &  NN & \proposed & NN & \proposed \\ \midrule
Hit@10 & 0.1344 & 0.3577 & \phantom{0.0}-\phantom{00} & 0.3496 \\
Hit@20 & 0.1999 & 0.3801 & \phantom{0.0}-\phantom{00} & 0.3690 \\ \bottomrule
    \end{tabular}%
    }
    \label{tab:4_4}
    \vspace{-2ex}
\end{table}
\subsubsection{Cross-Modal Retrieval via Modality Generation} \label{sec4_2_4}
{In this section, we demonstrate the potential of \proposed's generation-based approach for real-world industry applications by demonstrating its ability to retrieve items with missing modalities—something that conventional MRSs cannot achieve. For comparisons, we employed a nearest-neighbor (NN) approach since existing MRSs are {completely inapplicable} for this task. Specifically, the NN approach retrieves the top-5 similar items based on available modalities while utilizing the mean of the existing modality features for retrieval.}

{Table \ref{tab:4_4} presents the retrieval performance for cases with one missing modality and scenarios with two missing modalities (i.e., all modalities missing). \proposed~ significantly outperforms NN methods when a single modality is missing. Moreover, even in the challenging case where all modalities are missing, \proposed~ achieves remarkable performance, whereas NN fails entirely.}

We would like to emphasize that by leveraging generated modality features, \proposed~enables the retrieval of similar items and provides meaningful descriptions, even when modalities are missing during the item-streaming process. This highlights \proposed's strong practical applicability in real-world and industrial scenarios, where delivering effective recommendations and descriptions despite incomplete modality data is crucial.
\begin{table}[h]
\centering
\vspace{-2mm}
\caption{Ablation studies on the components of \proposed.}
\vspace{-3mm}
\resizebox{0.9\columnwidth}{!}{%
\begin{tabular}{l|cc|cc|cc}
\toprule
\textbf{Datasets} & \multicolumn{2}{c|}{\textbf{Baby}} & \multicolumn{2}{c|}{\textbf{Clothing}} & \multicolumn{2}{c}{\textbf{TikTok}} \\ \midrule
\textbf{Metric}      & R@20      & N@20      & R@20      & N@20          & R@20      & N@20    \\ \midrule
\proposed~           & \textbf{0.0897}    & \textbf{0.0404}    & \textbf{0.0725}    & \textbf{0.0324}        & \textbf{0.1093}    & \textbf{0.0476}  \\ \midrule
w/o Disentangle      & 0.0756    & 0.0331    & 0.0596    & 0.0268        & 0.0985    & 0.0419  \\
\quad w/o CLUB       & 0.0854    & 0.0373    & 0.0617    & 0.0277        & 0.1031    & 0.0452  \\
\quad w/o InfoNCE    & 0.0778    & 0.0347    & 0.0631    & 0.0280        & 0.1001    & 0.0429  \\  \midrule
w/o Generation       & 0.0848    & 0.0373    & 0.0646    & 0.0282        & 0.0988    & 0.0402  \\ 
\quad w/o Recon Loss & 0.0872    & 0.0376    & 0.0703    & 0.0313        & 0.1034    & 0.0434  \\
\quad w/o Gen Loss   & 0.0862    & 0.0374    & 0.0647    & 0.0284        & 0.1041    & 0.0438  \\ \midrule
w/o Alignment        & 0.0554    & 0.0248    & 0.0392    & 0.0142        & 0.0745    & 0.0335  \\
\quad w/o UI-align   & 0.0789    & 0.0335    & 0.0634    & 0.0284        & 0.0903    & 0.0389\\
\quad w/o BM-align   & 0.0811    & 0.0346    & 0.0576    & 0.0260        & 0.1011    & 0.0422 \\
\bottomrule
\end{tabular}%
}
\label{tab:4_5} 
\vspace{-2ex}
\end{table}
\subsection{Model Analysis}
\subsubsection{Ablation Study}
In Table~\ref{tab:4_5}, we conducted ablation studies to highlight the contribution of each component in \proposed. In general, excluding any loss resulted in a performance decline across all datasets. More precisely,
\textbf{1)} excluding \mm{\mc{L}_{disentangle}} led to a performance drop, demonstrating the effectiveness of disentangling modality features. This implies that modality disentanglement is effective for capturing modality representation.
\textbf{2)} Removing generation-related losses such as \mm{\mc{L}_{recon}} and \mm{\mc{L}_{gen}} resulted in performance declines. {These losses were introduced to effectively generate missing modalities, indicating that aligning reconstructed features with original features is crucial.}
\textbf{3)} {The removal of \mm{\mc{L}_{align}} led to the most significant performance degradation.}
This emphasizes the critical role of aligning modality features with the CF knowledge. Alignment with the CF knowledge ensures that modality features contribute effectively to the recommendation, demonstrating its importance in \proposed.

\begin{figure}[h!]
    \centering
    \vspace{-3mm}
    \includegraphics[width=0.9\linewidth]{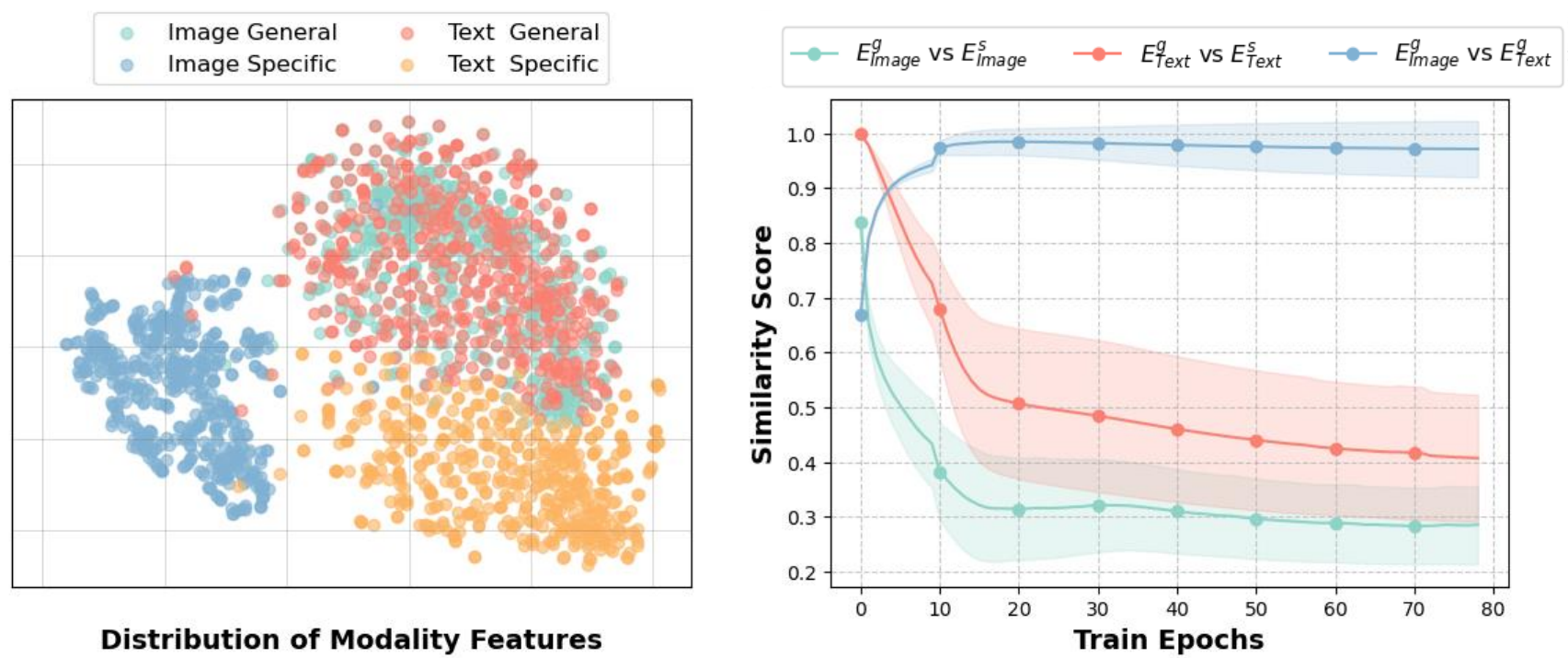} 
    \vspace{-2mm}
    \caption{(a) Visualization of disentangled modality features and (b) similarity score between the features during training} 
    \label{fig:4_6} 
    \vspace{-3mm}
\end{figure}
\subsubsection{Impact of Modality Disentanglement}
To evaluate the effectiveness of the Disentangling Modality Feature module, we visualized the general and specific modality features of 500 randomly selected items from the Baby dataset using TSNE \cite{van2008tsne} and tracked their similarity scores during training to observe the disentanglement process.
As shown in Figure \ref{fig:4_6}(a), specific features are well-separated, while general features appear more entangled, reflecting their shared nature across modalities.
Additionally, Figure \ref{fig:4_6}(b) shows that the similarity scores between general and specific features within the same modality gradually decrease during training, while the similarity scores between general features across different modalities increase.
These trends demonstrate the effectiveness of the disentanglement module in \proposed. 

\subsubsection{Time Complexity Analysis}
\begin{table}[h]
\centering
\vspace{-4mm}
\caption{Time Complexity of compared methods.}
\vspace{-3mm}
\resizebox{0.9\columnwidth}{!}{%
\begin{tabular}{c|ccc|ccc|ccc}
\toprule
Dataset & \multicolumn{3}{c|}{Baby} & \multicolumn{3}{c|}{Sports} & \multicolumn{3}{c}{Clothing} \\ \midrule
 (sec)   & Train & Inference & Total & Train & Inference & Total & Train & Inference & Total \\ \midrule
LGMRec & 2.81 & 3.39 & 171.5 &  7.01 & 6.54 &  603.2 &  6.58 & 7.20 &  559.1 \\ 
DAMRS  & 3.56 & 3.33 & 238.7 & 14.99 & 6.93 & 1723.6 & 16.53 & 7.23 & 1934.2 \\ 
GUME   & 2.74 & 2.94 & 152.1 &  9.45 & 7.32 &  662.1 &  9.29 & 7.96 &  667.2 \\ \midrule
\proposed & 2.63 & 3.38 & 210.4 &  8.83 & 6.18 &  666.5 &  9.91 & 7.32 &  802.4 \\ 
\bottomrule
\end{tabular}%
}
\label{tab:4_7} 
\vspace{-2mm}
\end{table}
The missing modality generation process in \proposed~ can incur additional computational overhead per specific epoch.
To evaluate the complexity of this process, we compared three key metrics: \textbf{1)} the average time per training epoch (Train), \textbf{2)} the total inference time (Inference), and \textbf{3)} the overall training time throughout the entire process (Total) across three Amazon datasets in Table \ref{tab:4_7}.

We have the following observations:
\textbf{1)} For the relatively small Baby dataset, the time required by \proposed~ is nearly identical to that of other models, demonstrating that the additional processes in \proposed~ do not create noticeable inefficiencies in smaller datasets.
\textbf{2)} In the larger Sports and Clothing datasets, while \proposed~incurs slightly higher overhead due to the modality generation and graph refinement processes, the difference remains modest.
Considering the significant performance gains achieved by \proposed, we argue that this additional time cost is reasonable, indicating that \proposed~ remains competitive even in larger-scale datasets.

\vspace{-2mm}
\begin{figure}[h!]
    \centering
    \includegraphics[width=0.9\linewidth]{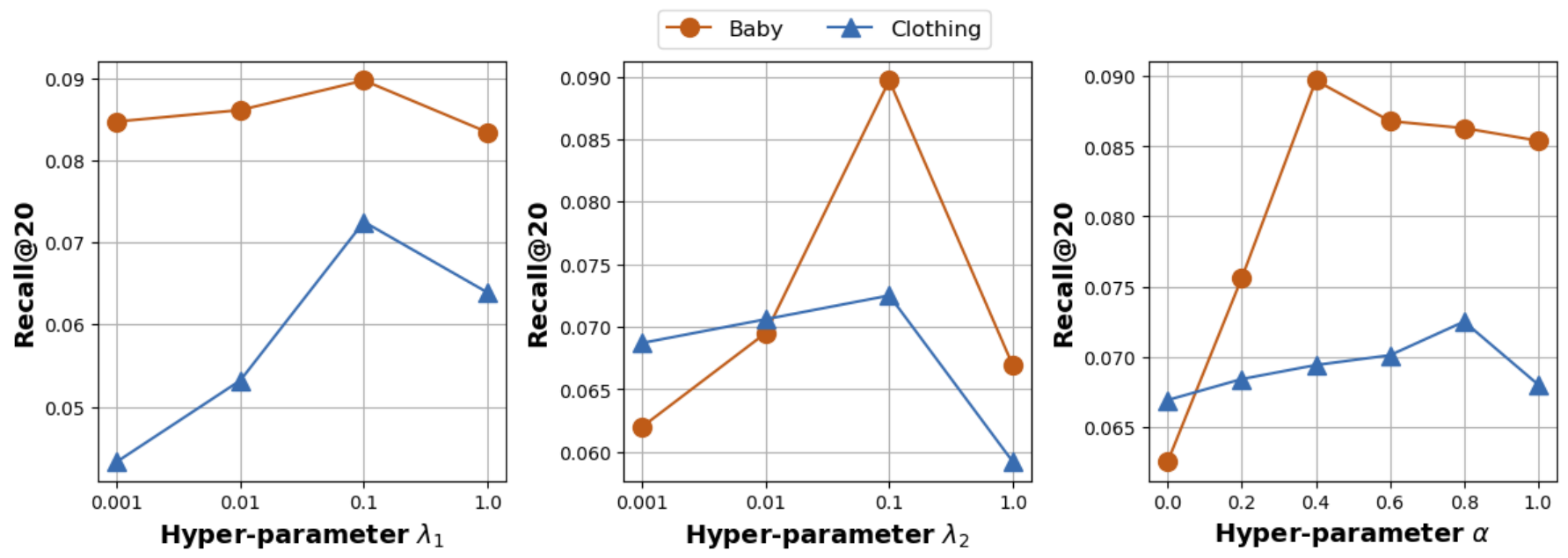} 
    \vspace{-3mm}
    \caption{Performance on different hyper-parameters} 
    \label{fig:4_3_4} 
    \vspace{-3mm}
\end{figure}

\subsubsection{Parameter Sensitivity}
In Figure \ref{fig:4_3_4}, we investigate the impact of \textit{every} hyperparameter in~\proposed, i.e., \mm{\lambda_1} for \mm{\mc{L}_{disentangle}}, \mm{\lambda_2} for \mm{\mc{L}_{align}}, and \mm{\alpha} for graph refinement, on the Baby and Clothing datasets.
\textbf{1)} The results show that both \mm{\lambda_1} and \mm{\lambda_2} achieve the best performance when set to 0.01, with significant performance drops observed when either value is too low or too high. Notably, \mm{\lambda_2}, which directly affects performance through alignment loss, is found to be more sensitive than \mm{\lambda_1}.
\textbf{2)} For the balancing hyperparameter \mm{\alpha}, a value of 0.0—indicating no graph refinement—leads to the lowest performance, emphasizing the importance of the graph refinement process in \proposed. Moreover, the best results are achieved with moderate \mm{\alpha} values, rather than extremes like 1.0, underscoring the need to maintain a balanced approach.

\section{Conclusion}
In this paper, we propose a novel model, \proposed~ that addresses two key challenges: 
{\textbf{1)} Missing modality scenarios are not sufficiently addressed, and 
\textbf{2)} Unique characteristics of modalities are overlooked.}
The core idea of \proposed~ lies in effectively disentangling general and specific modality features, which are then utilized to generate missing modalities in a fine-grained manner. 
These two modules work synergistically in \proposed~ framework. 
As a result, \proposed~ achieves high performance in realistic settings thus demonstrating its strong potential for industrial applications, including information retrieval tasks and scenarios with extremely missing modalities.

\subsection*{Acknowledgment} This work was supported by Institute of Information \& Communications Technology Planning \& Evaluation(IITP) grant funded by the Korea government(MSIT) (RS-2023-00216011, RS-2022-II220077), and National Research Foundation of Korea(NRF) funded by Ministry of Science and ICT (NRF-2022M3J6A1063021).
\bibliographystyle{ACM-Reference-Format}
\balance
\bibliography{acmart}


\begin{thebibliography}{40}


\ifx \showCODEN    \undefined \def \showCODEN     #1{\unskip}     \fi
\ifx \showDOI      \undefined \def \showDOI       #1{#1}\fi
\ifx \showISBNx    \undefined \def \showISBNx     #1{\unskip}     \fi
\ifx \showISBNxiii \undefined \def \showISBNxiii  #1{\unskip}     \fi
\ifx \showISSN     \undefined \def \showISSN      #1{\unskip}     \fi
\ifx \showLCCN     \undefined \def \showLCCN      #1{\unskip}     \fi
\ifx \shownote     \undefined \def \shownote      #1{#1}          \fi
\ifx \showarticletitle \undefined \def \showarticletitle #1{#1}   \fi
\ifx \showURL      \undefined \def \showURL       {\relax}        \fi
\providecommand\bibfield[2]{#2}
\providecommand\bibinfo[2]{#2}
\providecommand\natexlab[1]{#1}
\providecommand\showeprint[2][]{arXiv:#2}

\bibitem[Bai et~al\mbox{.}(2024)]%
        {bai2024milk}
\bibfield{author}{\bibinfo{person}{Haoyue Bai}, \bibinfo{person}{Le Wu}, \bibinfo{person}{Min Hou}, \bibinfo{person}{Miaomiao Cai}, \bibinfo{person}{Zhuangzhuang He}, \bibinfo{person}{Yuyang Zhou}, \bibinfo{person}{Richang Hong}, {and} \bibinfo{person}{Meng Wang}.} \bibinfo{year}{2024}\natexlab{}.
\newblock \showarticletitle{Multimodality invariant learning for multimedia-based new item recommendation}. In \bibinfo{booktitle}{\emph{Proceedings of the 47th International ACM SIGIR Conference on Research and Development in Information Retrieval}}. \bibinfo{pages}{677--686}.
\newblock


\bibitem[Cheng et~al\mbox{.}(2020)]%
        {cheng2020club}
\bibfield{author}{\bibinfo{person}{Pengyu Cheng}, \bibinfo{person}{Weituo Hao}, \bibinfo{person}{Shuyang Dai}, \bibinfo{person}{Jiachang Liu}, \bibinfo{person}{Zhe Gan}, {and} \bibinfo{person}{Lawrence Carin}.} \bibinfo{year}{2020}\natexlab{}.
\newblock \showarticletitle{Club: A contrastive log-ratio upper bound of mutual information}. In \bibinfo{booktitle}{\emph{International conference on machine learning}}. PMLR, \bibinfo{pages}{1779--1788}.
\newblock


\bibitem[Cho et~al\mbox{.}(2021)]%
        {cho2021missing}
\bibfield{author}{\bibinfo{person}{Jae~Won Cho}, \bibinfo{person}{Dong-Jin Kim}, \bibinfo{person}{Jinsoo Choi}, \bibinfo{person}{Yunjae Jung}, {and} \bibinfo{person}{In~So Kweon}.} \bibinfo{year}{2021}\natexlab{}.
\newblock \showarticletitle{Dealing with missing modalities in the visual question answer-difference prediction task through knowledge distillation}. In \bibinfo{booktitle}{\emph{Proceedings of the IEEE/CVF Conference on Computer Vision and Pattern Recognition}}. \bibinfo{pages}{1592--1601}.
\newblock


\bibitem[Diederik(2014)]%
        {diederik2014adam}
\bibfield{author}{\bibinfo{person}{P~Kingma Diederik}.} \bibinfo{year}{2014}\natexlab{}.
\newblock \showarticletitle{Adam: A method for stochastic optimization}.
\newblock \bibinfo{journal}{\emph{(No Title)}} (\bibinfo{year}{2014}).
\newblock


\bibitem[Ganh{\"o}r et~al\mbox{.}(2024)]%
        {ganhor2024sibrar}
\bibfield{author}{\bibinfo{person}{Christian Ganh{\"o}r}, \bibinfo{person}{Marta Moscati}, \bibinfo{person}{Anna Hausberger}, \bibinfo{person}{Shah Nawaz}, {and} \bibinfo{person}{Markus Schedl}.} \bibinfo{year}{2024}\natexlab{}.
\newblock \showarticletitle{A Multimodal Single-Branch Embedding Network for Recommendation in Cold-Start and Missing Modality Scenarios}. In \bibinfo{booktitle}{\emph{Proceedings of the 18th ACM Conference on Recommender Systems}}. \bibinfo{pages}{380--390}.
\newblock


\bibitem[Guo et~al\mbox{.}(2024)]%
        {guo2024lgmrec}
\bibfield{author}{\bibinfo{person}{Zhiqiang Guo}, \bibinfo{person}{Jianjun Li}, \bibinfo{person}{Guohui Li}, \bibinfo{person}{Chaoyang Wang}, \bibinfo{person}{Si Shi}, {and} \bibinfo{person}{Bin Ruan}.} \bibinfo{year}{2024}\natexlab{}.
\newblock \showarticletitle{LGMRec: Local and Global Graph Learning for Multimodal Recommendation}. In \bibinfo{booktitle}{\emph{Proceedings of the AAAI Conference on Artificial Intelligence}}, Vol.~\bibinfo{volume}{38}. \bibinfo{pages}{8454--8462}.
\newblock


\bibitem[He and McAuley(2016a)]%
        {he2016cnn}
\bibfield{author}{\bibinfo{person}{Ruining He} {and} \bibinfo{person}{Julian McAuley}.} \bibinfo{year}{2016}\natexlab{a}.
\newblock \showarticletitle{Ups and downs: Modeling the visual evolution of fashion trends with one-class collaborative filtering}. In \bibinfo{booktitle}{\emph{proceedings of the 25th international conference on world wide web}}. \bibinfo{pages}{507--517}.
\newblock


\bibitem[He and McAuley(2016b)]%
        {he2016vbpr}
\bibfield{author}{\bibinfo{person}{Ruining He} {and} \bibinfo{person}{Julian McAuley}.} \bibinfo{year}{2016}\natexlab{b}.
\newblock \showarticletitle{VBPR: visual bayesian personalized ranking from implicit feedback}. In \bibinfo{booktitle}{\emph{Proceedings of the AAAI conference on artificial intelligence}}, Vol.~\bibinfo{volume}{30}.
\newblock


\bibitem[He et~al\mbox{.}(2020)]%
        {he2020lightgcn}
\bibfield{author}{\bibinfo{person}{Xiangnan He}, \bibinfo{person}{Kuan Deng}, \bibinfo{person}{Xiang Wang}, \bibinfo{person}{Yan Li}, \bibinfo{person}{Yongdong Zhang}, {and} \bibinfo{person}{Meng Wang}.} \bibinfo{year}{2020}\natexlab{}.
\newblock \showarticletitle{Lightgcn: Simplifying and powering graph convolution network for recommendation}. In \bibinfo{booktitle}{\emph{Proceedings of the 43rd International ACM SIGIR conference on research and development in Information Retrieval}}. \bibinfo{pages}{639--648}.
\newblock


\bibitem[Kim et~al\mbox{.}(2024)]%
        {kim2024allmrec}
\bibfield{author}{\bibinfo{person}{Sein Kim}, \bibinfo{person}{Hongseok Kang}, \bibinfo{person}{Seungyoon Choi}, \bibinfo{person}{Donghyun Kim}, \bibinfo{person}{Minchul Yang}, {and} \bibinfo{person}{Chanyoung Park}.} \bibinfo{year}{2024}\natexlab{}.
\newblock \showarticletitle{Large language models meet collaborative filtering: An efficient all-round llm-based recommender system}. In \bibinfo{booktitle}{\emph{Proceedings of the 30th ACM SIGKDD Conference on Knowledge Discovery and Data Mining}}. \bibinfo{pages}{1395--1406}.
\newblock


\bibitem[Lin et~al\mbox{.}(2024)]%
        {lin2024gume}
\bibfield{author}{\bibinfo{person}{Guojiao Lin}, \bibinfo{person}{Meng Zhen}, \bibinfo{person}{Dongjie Wang}, \bibinfo{person}{Qingqing Long}, \bibinfo{person}{Yuanchun Zhou}, {and} \bibinfo{person}{Meng Xiao}.} \bibinfo{year}{2024}\natexlab{}.
\newblock \showarticletitle{GUME: Graphs and User Modalities Enhancement for Long-Tail Multimodal Recommendation}. In \bibinfo{booktitle}{\emph{Proceedings of the 33rd ACM International Conference on Information and Knowledge Management}}. \bibinfo{pages}{1400--1409}.
\newblock


\bibitem[Lin et~al\mbox{.}(2023)]%
        {lin2023ci2mg}
\bibfield{author}{\bibinfo{person}{Zhenghong Lin}, \bibinfo{person}{Yanchao Tan}, \bibinfo{person}{Yunfei Zhan}, \bibinfo{person}{Weiming Liu}, \bibinfo{person}{Fan Wang}, \bibinfo{person}{Chaochao Chen}, \bibinfo{person}{Shiping Wang}, {and} \bibinfo{person}{Carl Yang}.} \bibinfo{year}{2023}\natexlab{}.
\newblock \showarticletitle{Contrastive intra-and inter-modality generation for enhancing incomplete multimedia recommendation}. In \bibinfo{booktitle}{\emph{Proceedings of the 31st ACM International Conference on Multimedia}}. \bibinfo{pages}{6234--6242}.
\newblock


\bibitem[Malitesta et~al\mbox{.}(2024a)]%
        {malitesta2024dowedrop}
\bibfield{author}{\bibinfo{person}{Daniele Malitesta}, \bibinfo{person}{Emanuele Rossi}, \bibinfo{person}{Claudio Pomo}, \bibinfo{person}{Tommaso Di~Noia}, {and} \bibinfo{person}{Fragkiskos~D Malliaros}.} \bibinfo{year}{2024}\natexlab{a}.
\newblock \showarticletitle{Do We Really Need to Drop Items with Missing Modalities in Multimodal Recommendation?}. In \bibinfo{booktitle}{\emph{Proceedings of the 33rd ACM International Conference on Information and Knowledge Management}}. \bibinfo{pages}{3943--3948}.
\newblock


\bibitem[Malitesta et~al\mbox{.}(2024b)]%
        {malitesta2024missingpropar}
\bibfield{author}{\bibinfo{person}{Daniele Malitesta}, \bibinfo{person}{Emanuele Rossi}, \bibinfo{person}{Claudio Pomo}, \bibinfo{person}{Fragkiskos~D Malliaros}, {and} \bibinfo{person}{Tommaso Di~Noia}.} \bibinfo{year}{2024}\natexlab{b}.
\newblock \showarticletitle{Dealing with Missing Modalities in Multimodal Recommendation: a Feature Propagation-based Approach}.
\newblock \bibinfo{journal}{\emph{arXiv preprint arXiv:2403.19841}} (\bibinfo{year}{2024}).
\newblock


\bibitem[Oord et~al\mbox{.}(2018)]%
        {oord2018infonce}
\bibfield{author}{\bibinfo{person}{Aaron van~den Oord}, \bibinfo{person}{Yazhe Li}, {and} \bibinfo{person}{Oriol Vinyals}.} \bibinfo{year}{2018}\natexlab{}.
\newblock \showarticletitle{Representation learning with contrastive predictive coding}.
\newblock \bibinfo{journal}{\emph{arXiv preprint arXiv:1807.03748}} (\bibinfo{year}{2018}).
\newblock


\bibitem[Paszke et~al\mbox{.}(2019)]%
        {paszke2019pytorch}
\bibfield{author}{\bibinfo{person}{Adam Paszke}, \bibinfo{person}{Sam Gross}, \bibinfo{person}{Francisco Massa}, \bibinfo{person}{Adam Lerer}, \bibinfo{person}{James Bradbury}, \bibinfo{person}{Gregory Chanan}, \bibinfo{person}{Trevor Killeen}, \bibinfo{person}{Zeming Lin}, \bibinfo{person}{Natalia Gimelshein}, \bibinfo{person}{Luca Antiga}, {et~al\mbox{.}}} \bibinfo{year}{2019}\natexlab{}.
\newblock \showarticletitle{Pytorch: An imperative style, high-performance deep learning library}.
\newblock \bibinfo{journal}{\emph{Advances in neural information processing systems}}  \bibinfo{volume}{32} (\bibinfo{year}{2019}).
\newblock


\bibitem[Reimers(2019)]%
        {reimers2019sentenceBert}
\bibfield{author}{\bibinfo{person}{N Reimers}.} \bibinfo{year}{2019}\natexlab{}.
\newblock \showarticletitle{Sentence-BERT: Sentence Embeddings using Siamese BERT-Networks}.
\newblock \bibinfo{journal}{\emph{arXiv preprint arXiv:1908.10084}} (\bibinfo{year}{2019}).
\newblock


\bibitem[Rendle et~al\mbox{.}(2012)]%
        {rendle2012bpr}
\bibfield{author}{\bibinfo{person}{Steffen Rendle}, \bibinfo{person}{Christoph Freudenthaler}, \bibinfo{person}{Zeno Gantner}, {and} \bibinfo{person}{Lars Schmidt-Thieme}.} \bibinfo{year}{2012}\natexlab{}.
\newblock \showarticletitle{BPR: Bayesian personalized ranking from implicit feedback}.
\newblock \bibinfo{journal}{\emph{arXiv preprint arXiv:1205.2618}} (\bibinfo{year}{2012}).
\newblock


\bibitem[Tao et~al\mbox{.}(2022)]%
        {tao2022slmrec}
\bibfield{author}{\bibinfo{person}{Zhulin Tao}, \bibinfo{person}{Xiaohao Liu}, \bibinfo{person}{Yewei Xia}, \bibinfo{person}{Xiang Wang}, \bibinfo{person}{Lifang Yang}, \bibinfo{person}{Xianglin Huang}, {and} \bibinfo{person}{Tat-Seng Chua}.} \bibinfo{year}{2022}\natexlab{}.
\newblock \showarticletitle{Self-supervised learning for multimedia recommendation}.
\newblock \bibinfo{journal}{\emph{IEEE Transactions on Multimedia}}  \bibinfo{volume}{25} (\bibinfo{year}{2022}), \bibinfo{pages}{5107--5116}.
\newblock


\bibitem[Van~der Maaten and Hinton(2008)]%
        {van2008tsne}
\bibfield{author}{\bibinfo{person}{Laurens Van~der Maaten} {and} \bibinfo{person}{Geoffrey Hinton}.} \bibinfo{year}{2008}\natexlab{}.
\newblock \showarticletitle{Visualizing data using t-SNE.}
\newblock \bibinfo{journal}{\emph{Journal of machine learning research}} \bibinfo{volume}{9}, \bibinfo{number}{11} (\bibinfo{year}{2008}).
\newblock


\bibitem[Wang et~al\mbox{.}(2018)]%
        {wang2018lrmm}
\bibfield{author}{\bibinfo{person}{Cheng Wang}, \bibinfo{person}{Mathias Niepert}, {and} \bibinfo{person}{Hui Li}.} \bibinfo{year}{2018}\natexlab{}.
\newblock \showarticletitle{LRMM: Learning to recommend with missing modalities}.
\newblock \bibinfo{journal}{\emph{arXiv preprint arXiv:1808.06791}} (\bibinfo{year}{2018}).
\newblock


\bibitem[Wang et~al\mbox{.}(2019)]%
        {wang2019ngcf}
\bibfield{author}{\bibinfo{person}{Xiang Wang}, \bibinfo{person}{Xiangnan He}, \bibinfo{person}{Meng Wang}, \bibinfo{person}{Fuli Feng}, {and} \bibinfo{person}{Tat-Seng Chua}.} \bibinfo{year}{2019}\natexlab{}.
\newblock \showarticletitle{Neural graph collaborative filtering}. In \bibinfo{booktitle}{\emph{Proceedings of the 42nd international ACM SIGIR conference on Research and development in Information Retrieval}}. \bibinfo{pages}{165--174}.
\newblock


\bibitem[Wei et~al\mbox{.}(2023)]%
        {wei2023mmssl}
\bibfield{author}{\bibinfo{person}{Wei Wei}, \bibinfo{person}{Chao Huang}, \bibinfo{person}{Lianghao Xia}, {and} \bibinfo{person}{Chuxu Zhang}.} \bibinfo{year}{2023}\natexlab{}.
\newblock \showarticletitle{Multi-modal self-supervised learning for recommendation}. In \bibinfo{booktitle}{\emph{Proceedings of the ACM Web Conference 2023}}. \bibinfo{pages}{790--800}.
\newblock


\bibitem[Wei et~al\mbox{.}(2020)]%
        {wei2020grcn}
\bibfield{author}{\bibinfo{person}{Yinwei Wei}, \bibinfo{person}{Xiang Wang}, \bibinfo{person}{Liqiang Nie}, \bibinfo{person}{Xiangnan He}, {and} \bibinfo{person}{Tat-Seng Chua}.} \bibinfo{year}{2020}\natexlab{}.
\newblock \showarticletitle{Graph-refined convolutional network for multimedia recommendation with implicit feedback}. In \bibinfo{booktitle}{\emph{Proceedings of the 28th ACM international conference on multimedia}}. \bibinfo{pages}{3541--3549}.
\newblock


\bibitem[Wei et~al\mbox{.}(2019)]%
        {wei2019mmgcn}
\bibfield{author}{\bibinfo{person}{Yinwei Wei}, \bibinfo{person}{Xiang Wang}, \bibinfo{person}{Liqiang Nie}, \bibinfo{person}{Xiangnan He}, \bibinfo{person}{Richang Hong}, {and} \bibinfo{person}{Tat-Seng Chua}.} \bibinfo{year}{2019}\natexlab{}.
\newblock \showarticletitle{MMGCN: Multi-modal graph convolution network for personalized recommendation of micro-video}. In \bibinfo{booktitle}{\emph{Proceedings of the 27th ACM international conference on multimedia}}. \bibinfo{pages}{1437--1445}.
\newblock


\bibitem[Wu et~al\mbox{.}(2021)]%
        {wu2021sgl}
\bibfield{author}{\bibinfo{person}{Jiancan Wu}, \bibinfo{person}{Xiang Wang}, \bibinfo{person}{Fuli Feng}, \bibinfo{person}{Xiangnan He}, \bibinfo{person}{Liang Chen}, \bibinfo{person}{Jianxun Lian}, {and} \bibinfo{person}{Xing Xie}.} \bibinfo{year}{2021}\natexlab{}.
\newblock \showarticletitle{Self-supervised graph learning for recommendation}. In \bibinfo{booktitle}{\emph{Proceedings of the 44th international ACM SIGIR conference on research and development in information retrieval}}. \bibinfo{pages}{726--735}.
\newblock


\bibitem[Wu et~al\mbox{.}(2020)]%
        {wu2020incomplete2}
\bibfield{author}{\bibinfo{person}{Le Wu}, \bibinfo{person}{Yonghui Yang}, \bibinfo{person}{Kun Zhang}, \bibinfo{person}{Richang Hong}, \bibinfo{person}{Yanjie Fu}, {and} \bibinfo{person}{Meng Wang}.} \bibinfo{year}{2020}\natexlab{}.
\newblock \showarticletitle{Joint item recommendation and attribute inference: An adaptive graph convolutional network approach}. In \bibinfo{booktitle}{\emph{Proceedings of the 43rd International ACM SIGIR conference on research and development in Information Retrieval}}. \bibinfo{pages}{679--688}.
\newblock


\bibitem[Wu et~al\mbox{.}(2024)]%
        {wu2024surveyMissing}
\bibfield{author}{\bibinfo{person}{Renjie Wu}, \bibinfo{person}{Hu Wang}, \bibinfo{person}{Hsiang-Ting Chen}, {and} \bibinfo{person}{Gustavo Carneiro}.} \bibinfo{year}{2024}\natexlab{}.
\newblock \showarticletitle{Deep multimodal learning with missing modality: A survey}.
\newblock \bibinfo{journal}{\emph{arXiv preprint arXiv:2409.07825}} (\bibinfo{year}{2024}).
\newblock


\bibitem[Xv et~al\mbox{.}(2024)]%
        {xv2024damrs}
\bibfield{author}{\bibinfo{person}{Guipeng Xv}, \bibinfo{person}{Xinyu Li}, \bibinfo{person}{Ruobing Xie}, \bibinfo{person}{Chen Lin}, \bibinfo{person}{Chong Liu}, \bibinfo{person}{Feng Xia}, \bibinfo{person}{Zhanhui Kang}, {and} \bibinfo{person}{Leyu Lin}.} \bibinfo{year}{2024}\natexlab{}.
\newblock \showarticletitle{Improving Multi-modal Recommender Systems by Denoising and Aligning Multi-modal Content and User Feedback}. In \bibinfo{booktitle}{\emph{Proceedings of the 30th ACM SIGKDD Conference on Knowledge Discovery and Data Mining}}. \bibinfo{pages}{3645--3656}.
\newblock


\bibitem[Yu et~al\mbox{.}(2022)]%
        {yu2022simgcl}
\bibfield{author}{\bibinfo{person}{Junliang Yu}, \bibinfo{person}{Hongzhi Yin}, \bibinfo{person}{Xin Xia}, \bibinfo{person}{Tong Chen}, \bibinfo{person}{Lizhen Cui}, {and} \bibinfo{person}{Quoc Viet~Hung Nguyen}.} \bibinfo{year}{2022}\natexlab{}.
\newblock \showarticletitle{Are graph augmentations necessary? simple graph contrastive learning for recommendation}. In \bibinfo{booktitle}{\emph{Proceedings of the 45th international ACM SIGIR conference on research and development in information retrieval}}. \bibinfo{pages}{1294--1303}.
\newblock


\bibitem[Yu et~al\mbox{.}(2023)]%
        {yu2023mgcn}
\bibfield{author}{\bibinfo{person}{Penghang Yu}, \bibinfo{person}{Zhiyi Tan}, \bibinfo{person}{Guanming Lu}, {and} \bibinfo{person}{Bing-Kun Bao}.} \bibinfo{year}{2023}\natexlab{}.
\newblock \showarticletitle{Multi-view graph convolutional network for multimedia recommendation}. In \bibinfo{booktitle}{\emph{Proceedings of the 31st ACM International Conference on Multimedia}}. \bibinfo{pages}{6576--6585}.
\newblock


\bibitem[Yu et~al\mbox{.}(2021)]%
        {yu2021modalspecificMSA}
\bibfield{author}{\bibinfo{person}{Wenmeng Yu}, \bibinfo{person}{Hua Xu}, \bibinfo{person}{Ziqi Yuan}, {and} \bibinfo{person}{Jiele Wu}.} \bibinfo{year}{2021}\natexlab{}.
\newblock \showarticletitle{Learning modality-specific representations with self-supervised multi-task learning for multimodal sentiment analysis}. In \bibinfo{booktitle}{\emph{Proceedings of the AAAI conference on artificial intelligence}}, Vol.~\bibinfo{volume}{35}. \bibinfo{pages}{10790--10797}.
\newblock


\bibitem[Yuan et~al\mbox{.}(2023)]%
        {yuan2023wheretogo}
\bibfield{author}{\bibinfo{person}{Zheng Yuan}, \bibinfo{person}{Fajie Yuan}, \bibinfo{person}{Yu Song}, \bibinfo{person}{Youhua Li}, \bibinfo{person}{Junchen Fu}, \bibinfo{person}{Fei Yang}, \bibinfo{person}{Yunzhu Pan}, {and} \bibinfo{person}{Yongxin Ni}.} \bibinfo{year}{2023}\natexlab{}.
\newblock \showarticletitle{Where to go next for recommender systems? id-vs. modality-based recommender models revisited}. In \bibinfo{booktitle}{\emph{Proceedings of the 46th International ACM SIGIR Conference on Research and Development in Information Retrieval}}. \bibinfo{pages}{2639--2649}.
\newblock


\bibitem[Zhang et~al\mbox{.}(2021)]%
        {zhang2021lattice}
\bibfield{author}{\bibinfo{person}{Jinghao Zhang}, \bibinfo{person}{Yanqiao Zhu}, \bibinfo{person}{Qiang Liu}, \bibinfo{person}{Shu Wu}, \bibinfo{person}{Shuhui Wang}, {and} \bibinfo{person}{Liang Wang}.} \bibinfo{year}{2021}\natexlab{}.
\newblock \showarticletitle{Mining latent structures for multimedia recommendation}. In \bibinfo{booktitle}{\emph{Proceedings of the 29th ACM international conference on multimedia}}. \bibinfo{pages}{3872--3880}.
\newblock


\bibitem[Zhang et~al\mbox{.}(2022)]%
        {zhang2022micro}
\bibfield{author}{\bibinfo{person}{Jinghao Zhang}, \bibinfo{person}{Yanqiao Zhu}, \bibinfo{person}{Qiang Liu}, \bibinfo{person}{Mengqi Zhang}, \bibinfo{person}{Shu Wu}, {and} \bibinfo{person}{Liang Wang}.} \bibinfo{year}{2022}\natexlab{}.
\newblock \showarticletitle{Latent structure mining with contrastive modality fusion for multimedia recommendation}.
\newblock \bibinfo{journal}{\emph{IEEE Transactions on Knowledge and Data Engineering}} \bibinfo{volume}{35}, \bibinfo{number}{9} (\bibinfo{year}{2022}), \bibinfo{pages}{9154--9167}.
\newblock


\bibitem[Zhou et~al\mbox{.}(2023b)]%
        {zhou2023mmsurvey}
\bibfield{author}{\bibinfo{person}{Hongyu Zhou}, \bibinfo{person}{Xin Zhou}, \bibinfo{person}{Zhiwei Zeng}, \bibinfo{person}{Lingzi Zhang}, {and} \bibinfo{person}{Zhiqi Shen}.} \bibinfo{year}{2023}\natexlab{b}.
\newblock \showarticletitle{A comprehensive survey on multimodal recommender systems: Taxonomy, evaluation, and future directions}.
\newblock \bibinfo{journal}{\emph{arXiv preprint arXiv:2302.04473}} (\bibinfo{year}{2023}).
\newblock


\bibitem[Zhou(2023)]%
        {zhou2023mmrec}
\bibfield{author}{\bibinfo{person}{Xin Zhou}.} \bibinfo{year}{2023}\natexlab{}.
\newblock \showarticletitle{Mmrec: Simplifying multimodal recommendation}. In \bibinfo{booktitle}{\emph{Proceedings of the 5th ACM International Conference on Multimedia in Asia Workshops}}. \bibinfo{pages}{1--2}.
\newblock


\bibitem[Zhou and Shen(2023)]%
        {zhou2023freedom}
\bibfield{author}{\bibinfo{person}{Xin Zhou} {and} \bibinfo{person}{Zhiqi Shen}.} \bibinfo{year}{2023}\natexlab{}.
\newblock \showarticletitle{A tale of two graphs: Freezing and denoising graph structures for multimodal recommendation}. In \bibinfo{booktitle}{\emph{Proceedings of the 31st ACM International Conference on Multimedia}}. \bibinfo{pages}{935--943}.
\newblock


\bibitem[Zhou et~al\mbox{.}(2023a)]%
        {zhou2023bm3}
\bibfield{author}{\bibinfo{person}{Xin Zhou}, \bibinfo{person}{Hongyu Zhou}, \bibinfo{person}{Yong Liu}, \bibinfo{person}{Zhiwei Zeng}, \bibinfo{person}{Chunyan Miao}, \bibinfo{person}{Pengwei Wang}, \bibinfo{person}{Yuan You}, {and} \bibinfo{person}{Feijun Jiang}.} \bibinfo{year}{2023}\natexlab{a}.
\newblock \showarticletitle{Bootstrap latent representations for multi-modal recommendation}. In \bibinfo{booktitle}{\emph{Proceedings of the ACM Web Conference 2023}}. \bibinfo{pages}{845--854}.
\newblock


\bibitem[Zhu and Hill(2021)]%
        {zhu2021incomplete1}
\bibfield{author}{\bibinfo{person}{Lipeng Zhu} {and} \bibinfo{person}{David~J Hill}.} \bibinfo{year}{2021}\natexlab{}.
\newblock \showarticletitle{Data/model jointly driven high-quality case generation for power system dynamic stability assessment}.
\newblock \bibinfo{journal}{\emph{IEEE Transactions on Industrial Informatics}} \bibinfo{volume}{18}, \bibinfo{number}{8} (\bibinfo{year}{2021}), \bibinfo{pages}{5055--5066}.
\newblock


\end{thebibliography}
\end{document}